\documentclass[lettersize,journal]{IEEEtran}
\usepackage{amsmath,amsfonts}
\usepackage{algorithmic}
\usepackage{algorithm}
\usepackage{array}

\usepackage[caption=false,font=footnotesize,labelfont=rm,textfont=rm]{subfig}

\usepackage{textcomp}
\usepackage{stfloats}
\usepackage{url}
\usepackage{verbatim}
\usepackage{graphicx}
\usepackage{cite}
\usepackage{graphicx}

\usepackage{hyperref}  

\usepackage{tabularx}  

\usepackage{cleveref}

\usepackage{amsmath}  
\usepackage{multirow}

\usepackage{xspace}
\newcommand{\figrange}[2]{Fig.~\ref{#1}--\ref{#2}\xspace}

\usepackage{color}

\usepackage{array}

\usepackage{tablefootnote}

\usepackage{threeparttable}

\crefname{figure}{Fig.}{Figs.}
\Crefname{figure}{Fig.}{Figs.}
\begin{document}

\title{Unsupervised Learning-Based Joint Resource Allocation and Beamforming Design for RIS-Assisted MISO-OFDMA Systems}

\author{
	\IEEEauthorblockN{
		Yu Ma,
		Xingyu Zhou,
		\IEEEmembership{Graduate Student Member, IEEE}, 
		Xiao Li, 
		\IEEEmembership{Member, IEEE}, 
		Le Liang, 
		\IEEEmembership{Member, IEEE}, 
	    Shi Jin, 
		\IEEEmembership{Fellow, IEEE}
	}
	
	\thanks{Yu Ma, Xingyu Zhou, Xiao Li, and Shi Jin are with the National Mobile Communications Research Laboratory, Southeast University, Nanjing 210096, China (e-mail: yuma@seu.edu.cn; xy\_zhou@seu.edu.cn; li\_xiao@seu.edu.cn; jinshi@seu.edu.cn).
		
		Le Liang is with the National Mobile Communications Research Laboratory, Southeast University, Nanjing 210096, China, and also with the Purple Mountain Laboratories, Nanjing 211111, China (e-mail: lliang@seu.edu.cn).}
}

\markboth{Journal of \LaTeX\ Class Files,~Vol.~14, No.~8, August~2021}%
{Shell \MakeLowercase{\textit{et al.}}: A Sample Article Using IEEEtran.cls for IEEE Journals}


\maketitle

\begin{abstract}

Reconfigurable intelligent surface (RIS) is regarded as one of the pivotal technologies for sixth-generation wireless communication systems. Nevertheless, the inherent scarcity of wireless communication resources motivates the need for effective resource allocation schemes in RIS-assisted systems. This paper investigates the downlink transmission of an RIS-assisted multiple-input single-output (MISO) orthogonal frequency division multiple access (OFDMA) communication systems. To achieve a high system sum rate with low computational complexity, we develop a two-stage unsupervised learning based approach with customized loss function  for the RIS reflection phase shift design, active beamforming at base station (BS) and time-frequency resource block (RB) allocation. The proposed approach consists of two neural networks: BeamNet, which takes channel state information (CSI) as input to predict the RIS reflection phase shift, and AllocationNet, which generates RB allocation decisions based on the equivalent CSI from the BS to the users, where the equivalent CSI is obtained by combining the original CSI with the RIS reflection phase shifts predicted by BeamNet. The active beamforming is implemented using the maximum ratio transmission and water-filling algorithm. In order to incorporate the discrete constraints of RIS reflection phase shift and RB allocation decisions into the network while maintaining network differentiability, we introduce a quantization function and the Gumbel softmax trick into BeamNet and AllocationNet, respectively. Furthermore, a customized loss function and phased training strategy are devised to enhance training efficiency and address quality-of-service constraints. Simulation results demonstrate that the proposed approach achieves 99.93\% of the system sum rate of the successive convex approximation (SCA) method while requiring only 0.036\% of its runtime. Additionally, the method’s effectiveness and robustness are validated under different delay tap numbers, user distributions, and Rician factors, demonstrating its strong adaptability to different communication environments.
\end{abstract}


\section{Introduction}
\IEEEPARstart{N}{owadays}, with the large-scale deployment of fifth-generation wireless communication systems (5G), the focus of research has gradually shifted to sixth-generation wireless communication systems (6G). Compared to 5G, 6G is expected to meet much higher performance requirements, such as enhanced spectrum and energy efficiency, higher peak and user-experienced data rates, increased area or spatial traffic capacity, greater connectivity density, lower latency, and improved mobility\cite{WirelessZhang}.
To meet these requirements, various enabling technologies have been proposed, such as ultra-massive multiple-input multiple-output (MIMO) \cite{UltraMassive}, terahertz (THz) communications \cite{Terahertz}, integrated sensing and communications (ISAC) \cite{ISAC1, ISAC2} and reconfigurable intelligent surfaces (RISs) \cite{reviewer2_1, reviewer2_2}.

As one of the key enabling technologies, RIS has emerged as a promising paradigm, and attracted significant attention from both the industry and the academia. It introduces a large number of low-cost, low-power programmable reflection elements into the communication environment, which can flexibly adjust the amplitude, phase, and other parameters of incident electromagnetic waves, thereby intelligently alter the propagation paths of wireless signals \cite{SmartDi}. Through proper configuration of the reflection elements, this novel technology can dynamically optimize wireless channel characteristics, so as to effectively enhance the signal quality, expand the coverage, and significantly reduce the energy consumption\cite{ReconfigurablePan, 9999288}. It is foreseeable that as relevant technologies continue to mature and improve, RIS will play a crucial role in next-generation wireless communication systems, providing strong support for the development of more efficient, intelligent, and sustainable communication networks\cite{ReconfigurableLiu}.

To explore the application of RIS in wireless communication systems, extensive efforts have been made in the design of RIS-assisted narrowband systems.
For instance, in early single-user scenarios, semi-definite relaxation (SDR)-based algorithms were applied to design active and passive beamforming in RIS-assisted multiple-input single-output (MISO) downlink systems \cite{single-user, single-user1}.
To extend these techniques to multi-user systems, Guo \emph{et al.} employed fractional programming combined with the alternating direction method of multipliers (ADMM) to maximize the weighted sum rate \cite{ADMM}, while other studies adopted SDR-based joint optimization methods \cite{SDR}.
However, these works commonly assumed continuous RIS reflection phase shifts.
To address practical hardware limitations, discrete phase shifts were considered in \cite{discrete, discrete1, discrete2}, where beamforming schemes were developed to strike a balance between performance and implementation feasibility.

Above works focused on narrowband frequency-flat fading channels.
In contrast, for more general frequency-selective channels, RIS reflection coefficients must adapt to multiple signal paths with different delays, making the optimization problem significantly more complex.
To address this, Hassouna \emph{et al.} proposed an iterative power allocation method combined with SDR to optimize RIS reflection phase shifts in RIS-assisted orthogonal frequency division multiplexing (OFDM) systems \cite{Rate}, while Feng \emph{et al.} employed an alternating optimization (AO) approach for joint base station (BS) beamforming and RIS phase design in RIS-assisted MISO-OFDM systems \cite{power2}.
Despite their effectiveness, both \cite{Rate} and \cite{power2} did not consider spectrum resource allocation, which is essential in wireless communication systems for efficient use of spectrum resource.
To bridge this gap, a harmony search-based AO algorithm was proposed in \cite{ofdmaLee} to jointly optimize subcarrier assignment, RIS relection phase shifts, and BS beamforming in RIS-assisted MISO-orthogonal frequency division multiple access (OFDMA) systems. 
Building on this direction, Gao \emph{et al.} investigated the integration of unmanned aerial vehicles (UAVs) with RIS-assisted OFDMA systems.
They formulated a non-convex optimization problem to maximize the system sum rate by jointly designing UAV trajectory, RIS scheduling, and subcarrier allocation, while meeting heterogeneous quality-of-service (QoS) requirements \cite{ofdmaWei}.
Although these traditional algorithms achieve promising performance, their reliance on iterative optimization typically incurs high computational complexity and limits real-time applicability.

To alleviate computational complexity and reduce dependence on expert-designed processes, deep learning (DL)-based methods have been increasingly adopted in wireless communication systems \cite{Liang}.
In early studies, supervised learning was used for channel estimation \cite{channelestimation} and power allocation \cite{supervised}.
Supervised learning has also been applied in RIS-assisted systems, such as for estimating direct and cascaded channels \cite{9090876}, and for designing passive beamforming \cite{9370097}.
However, supervised learning typically requires large volumes of labeled data \cite{label}, which are difficult to obtain in practical systems and may be as costly to generate as solving the original optimization problem \cite{Unsupervised_Gao}.
To address the data labeling challenge, some studies have explored semi-supervised \cite{Semi-Supervised} and self-supervised learning \cite{Self-Supervised}.
Nevertheless, semi-supervised learning still relies partly on labeled data, and its performance is sensitive to the quality and quantity of unlabeled data.
Meanwhile, self-supervised learning often suffers from task design complexity and limited generalization ability.
To overcome these limitations, unsupervised learning (UL)-based approaches have gained increasing attention, offering a way to achieve efficient model training without labeled data.
For example, UL has been used to jointly perform antenna selection and hybrid beamforming in MIMO systems \cite{Unsupervised_Gao}.
In RIS-assisted systems, UL-based methods have been developed for passive beamforming \cite{UnsupervisedJiabao, UnsupervisedSong}, and an UL-based algorithm was also proposed for RIS-assisted ISAC systems, achieving both low complexity and high efficiency \cite{10533223}.

Inspired by the demand for achieving low computational complexity in resource allocation for RIS-assisted downlink MISO-OFDMA wireless systems, this paper investigates an RIS-assisted MISO-OFDMA downlink communication system. The main contributions of this work are summarized as follows.

\begin{itemize}
	\item We study an RIS-assisted MISO-OFDMA downlink system aiming to maximize the sum rate by jointly optimizing RIS reflection phase shifts, time-frequency resource block (RB) allocation, and active beamforming at BS. To overcome RIS inflexibility, distinct phase shifts are assigned per timeslot. An UL-based algorithm with a custom loss function is proposed to solve the non-convex problem while satisfying QoS constraints.
	
	\item To reduce model complexity, a two-stage neural network is designed: BeamNet predicts RIS phase shifts, and AllocationNet handles RB allocation. Active beamforming is derived based on their outputs.
	
	\item Due to the large number of parameters in the entire network, joint training from scratch is inefficient. Therefore, this paper proposes a phased training approach to optimize the entire network and enhance training efficiency.
		
	\item Extensive simulation results demonstrate that the proposed UL-based algorithm can achieve 99.93\% of the system sum rate of the successive convex approximation (SCA) method while consuming only 0.036\% of its runtime. Moreover, it satisfies the QoS constraints. Additionally, the algorithm exhibits strong adaptability across different environments, including varying delay tap numbers, user distributions, and Rician factors.
\end{itemize}

The rest of the paper is organized as follows: In Section \ref{section:System Model and Problem formulation}, the system model is introduced. At the end of this section, the optimization problem is formally defined. Section \ref{section:Unsupervised Learning Based Algorithm} presents the design of a two-stage network structure along with an UL-based optimization algorithm to solve the optimization problem. Section \ref{section:Numerical Results} presents numerical simulations and complexity analysis to evaluate the performance of the proposed method. Finally, conclusions are presented in Section \ref{section:Conclusion}.
The notations are listed in \Cref{tab0}.

\begin{table}[h]
	\begin{center}
		\caption{Notations.}
		\label{tab0}
		\begin{tabular}{| c | c |}
			\hline
			Symbol & Description \\
			\hline
			$x$, $\mathbf{x}$, $\mathbf{X}$ & Scalar variable, vector and matrix, respectively \\
			\hline
			$\mathbb{C}$, $\mathbb{R}$ & Sets of complex and real numbers \\
			\hline
			$(\cdot)^T$ & Transpose operation \\
			\hline
			$(\cdot)^*$ & Conjugate operation \\
			\hline
			$(\cdot)^H$ & Conjugate transpose operation \\
			\hline
			$\Re(\cdot)$ & Real part of a complex number \\
			\hline
			$\Im(\cdot)$ & Imaginary part of a complex number \\
			\hline
			$\|\cdot\|$ & $\ell_2$ norm of a vector \\
			\hline
			$\mathbb{E}[\cdot]$ & Expectation operator \\
			\hline
			$\text{diag}(\cdot)$ & Diagonal matrix with entries from the input vector \\
			\hline
			$\mathbf{U}(m,:)$ & \(m\)-th row of matrix \(\mathbf{U}\) \\
			\hline
			$\mathbf{U}(:,n)$ & \(n\)-th column of matrix \(\mathbf{U}\) \\
			\hline
			$\mathbf{U}(m,n)$ & \((m,n)\)-th entry of matrix \(\mathbf{U}\) \\
			\hline
			$\mathcal{CN}(\mu, \sigma^2)$ & 
			\begin{tabular}[c]{@{}p{5.8cm}@{}}Circularly symmetric complex Gaussian distribution with mean \(\mu\) and variance \(\sigma^2\)\end{tabular} \\
			\hline
			$(x)^+$ & $\max(x, 0)$ \\
			\hline
		\end{tabular}
	\end{center}
\end{table}

\begin{figure}[t]
	\centering
	\includegraphics[width=0.5\textwidth, keepaspectratio]{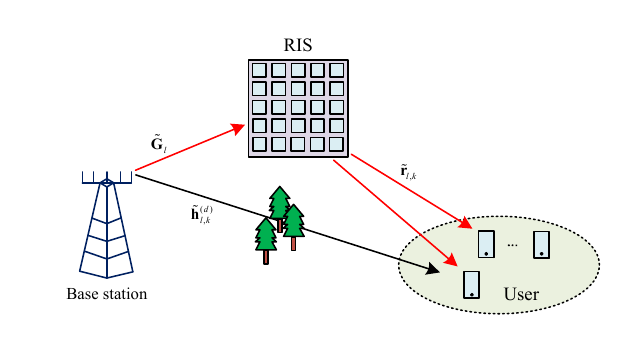}
	\caption{RIS-assisted downlink communication system.}
	\label{Fig:scenery}
\end{figure}

\section{System Model and Problem formulation}   \label{section:System Model and Problem formulation}

We consider an RIS-assisted MISO-OFDMA communication system as shown in \Cref{Fig:scenery}, where a BS equipped with $N_t$ antennas serves \( K \) single-antenna users. The RIS composed of \( M \) passive reflection elements is deployed to enhance the transmission effectively. Furthermore, we consider a quasi-static block fading channel model, whereby the channel remains constant within each coherence block. The overall system bandwidth is divided into $N$ subcarriers represented as $\mathcal{N}=\{1,2,\cdots,N\}$. On the other hand, the duration of each channel coherence block is divided into $Q$ equal-sized timeslots, denoted by the set $\mathcal{Q} = \{1, \dots, Q\}$.

\begin{figure}[t]
	\centering
	\includegraphics[width=0.5\textwidth, keepaspectratio]{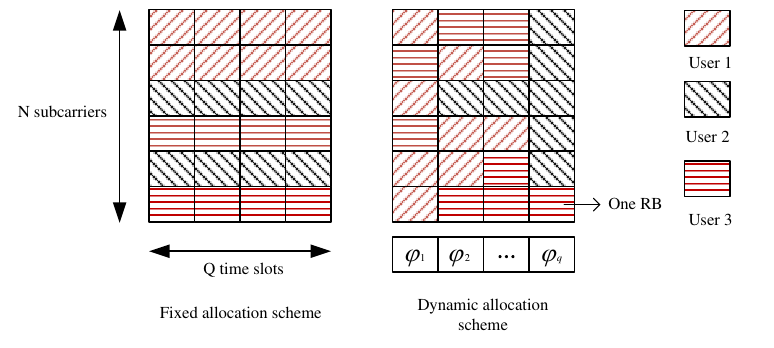}
	\caption{Illustration of fixed and dynamic allocation schemes: different fill patterns represent the allocation of RB to distinct users.}
	\label{Fig:passivebeamforming}
\end{figure}

Due to the lack of baseband signal processing capabilities in RIS, the RIS reflection phase shift at different subbands are the same under each RIS configuration. This limitation poses a significant challenge for RIS-assisted OFDMA systems, as the same reflection phase shift must be applied to all \(NK\) channels within each timeslot. When \(N\) and/or \(K\) are large, this configuration can lead to a significant degradation in system performance. To address this issue, we adopt a dynamic resource allocation scheme\cite{OFDMA_zhangRui}, which allocates the \(NQ\) time-frequency RBs within each channel coherence block across $K$ users, where each RB corresponds to a specific subcarrier $n$ in timeslot $q$. As illustrated in \Cref{Fig:passivebeamforming}, instead of employing the same RB allocation  and RIS reflection phase shift at each timeslot within the same channel coherence block, the dynamic resource allocation scheme could allocate different subcarriers to one user and adopt different RIS reflection phase shift at different timeslots. In particular, the dynamic resource allocation scheme is designed such that the corresponding optimal resource allocation allocates generally fewer users to be simultaneously served by the RIS at each time slot, thus reducing the number of channels that the reflection phase shift need to adapt to and thereby enhancing the RIS passive beamforming gain.

\subsection{Signal Model}

In the MISO-OFDMA system considered in this paper, $\tilde{\mathbf{h}}^{(d)}_{l,k}\in\mathbb{C}^{1\times N_t},l=0,1,...,L_0-1$ represents the time-domain baseband equivalent channel of the direct link from the BS to user \( k \), where $L_0 - 1$ denotes the number of tap delays. Similarly, $\mathbf{\tilde{G}}_l\in\mathbb{C}^{M\times N_t}, l=0,1,...,L_1 - 1$ and $\tilde{\mathbf{r}}_{l,k}\in\mathbb{C}^{1\times M},l=0,1,...,L_2 - 1$ respectively denote the time-domain baseband equivalent channels from the BS to the RIS and from the RIS to user $k$ link, where $L_1$ and $L_2$ are their respective numbers of tap delays. Therefore, the total maximum number of delay taps is $L=\operatorname*{max}\big\{L_{0},L_{1}+L_{2}-1\big\}$ \cite{singal}. Thus, the overall channel from the BS to user \( k \) at timeslot $q$ can be expressed as
\begin{equation}
	\begin{aligned}
		\tilde{\mathbf{h}}_{l,k,q}& =\tilde{\mathbf{h}}^{(d)}_{l,k}+\sum_{i=0}^{L_{2}-1}\tilde{\mathbf{r}}_{i,k}\boldsymbol{\Phi}_{q}\tilde{\mathbf{G}}_{l-i} \\
		&=\tilde{\mathbf{h}}^{(d)}_{l,k}+\sum_{i=0}^{L_{2}-1}\boldsymbol{\mathbf{\varphi}}_{q}^{T}\mathrm{diag}(\tilde{\mathbf{r}}_{i,k})\tilde{\mathbf{G}}_{l-i} \\
		&=\tilde{\mathbf{h}}^{(d)}_{l,k}+\boldsymbol{\mathbf{\varphi}}_{q}^{T}\sum_{i=0}^{L_{2}-1}\mathrm{diag}(\tilde{\mathbf{r}}_{i,k})\tilde{\mathbf{G}}_{l-i}, \\ 
		& \quad \quad \quad  \quad l=0,1,...,L -1,
	\end{aligned}
\end{equation}
where $\tilde{\mathbf{G}}_{l}=\mathbf{0} \text{,}\; l\in\begin{Bmatrix}1-L_{2}$\text{\,} $\ldots,-1\end{Bmatrix}\cup\begin{Bmatrix}L_{1},\ldots,L-1\end{Bmatrix}$, and $\boldsymbol{\Phi}_q=\operatorname{diag}(e^{j\theta_{1,q}},e^{j\theta_{2,q}},...,e^{j\theta_{M,q}})\in\mathbb{C}^{M\times M}$ is the  RIS reflection phase shift matrix at timeslot $q$. Let $\boldsymbol{\mathbf{\varphi}}_q=[e^{j\theta_{1,q}},e^{j\theta_{2,q}},...,e^{j\theta_{M,q}}]^T\in\mathbb{C}^{M\times1}$ denote the vector of RIS reflection phase shift at timeslot $q$, where $\theta_{m,q}$ refers to the phase shift of the $m$-th reflection element at timeslot $q$. We set $\tilde{\mathbf{h}}^{(r)}_{l,k} = \sum_{i=0}^{L_{2}-1}\mathrm{diag}(\tilde{\mathbf{r}}_{i,k})\tilde{\mathbf{G}}_{l-i}$, thus $\tilde{\mathbf{h}}_{l,k,q}$ can further be expressed as

\begin{equation}
	\tilde{\mathbf{h}}_{l,k,q}=\tilde{\mathbf{h}}^{(d)}_{l,k}+\boldsymbol{\mathbf{\varphi}}_{q}^{T}\tilde{\mathbf{h}}^{(r)}_{l,k}.
\end{equation}
To leverage the advantages of OFDM, we assume that the length of the cyclic prefix exceeds the maximum delay taps, i.e., \( N_{CP} \geq L \) \cite{prefix}. This ensures that inter-symbol interference is eliminated. Furthermore, the discrete Fourier transform (DFT) is applied to transform the time-domain channel into the frequency-domain channel \cite{DFT}, which can be expressed as

\begin{equation}
	\begin{aligned}
		\bar{\mathbf{h}}_{n,k,q}& =\sum_{l=0}^{L-1}\tilde{\mathbf{h}}_{l,k,q} e^{\frac{-j2\pi ln}{N}} \\
		&=\sum_{l=0}^{L_{0}-1}\tilde{\mathbf{h}}^{(d)}_{l,k} e^{\frac{-j2\pi ln}{N}}+\boldsymbol{\mathbf{\varphi}}_{q}^{T}\sum_{l=0}^{L_{1}+L_{2}-2}\tilde{\mathbf{h}}^{(r)}_{l,k} e^{\frac{-j2\pi ln}{N}} \\
		&=\bar{\mathbf{h}}^{(d)}_{n,k}+{\boldsymbol{\varphi}}_{q}^{T}\bar{\mathbf{h}}^{(r)}_{n,k} ,
	\end{aligned}
\end{equation}
where $\bar{\mathbf{h}}^{(d)}_{n,k}$ and $\bar{\mathbf{h}}^{(r)}_{n,k}$ represent the direct link channel and cascaded reflection channel, of user \( k \) on subcarrier \( n \) in the frequency domain, respectively. Therefore, in the  timeslot \( q \), the received signal of user \( k \) on subcarrier \( n \) can be represented as

\begin{equation}
	{ 
	y_{n,k,q}=(\bar{\mathbf{h}}^{(d)}_{n,k}+\boldsymbol{\varphi}_{q}^T\bar{\mathbf{h}}^{(r)}_{n,k})\mathbf{w}_{n,q}s_{n,q}+\nu_{n,q},} 
\end{equation}

\noindent where \( \mathbf{w}_{n,q} \in \mathbb{C}^{N_t \times 1} \) denotes the beamforming vector on subcarrier \( n \) in the timeslot \( q \), \( s_{n,q} \) represents the transmitted signal on subcarrier \( n \) in the timeslot \( q \) with \( \mathbb{E}\{| s_{n,q} |^2 \} = 1 \), and 
 \( \nu_{n,q} \) is additive Gaussian white noise, which satisfies \( \nu_{n,q} \sim \mathcal{CN}(0, \sigma^2)\).
Thus, in the  timeslot \( q \), the signal-to-noise ratio (SNR) of user \( k \) on subcarrier \( n \) can be expressed as

\begin{equation}
	\gamma_{n,k,q}=\frac{\left|(\bar{\mathbf{h}}^{(d)}_{n,k}+\boldsymbol{\varphi}_{{q}}^T\bar{\mathbf{h}}^{(r)}_{n,k})\mathbf{w}_{n,q}\right|^2}{\sigma^2}.
\end{equation}
Therefore, the achievable rate for user \( k \) is given by

\begin{equation}
	R_{k}=\frac{W}{Q}\sum_{q=1}^{Q}\sum_{n=1}^{N}\alpha_{n,k,q}\mathrm{log}_2(1+\gamma_{n,k,q}),
\end{equation}
where \( W \) represents the bandwidth of the subcarrier, and \( \alpha_{n,k,q} \) denotes the RB allocation decisions,  so that $\alpha_{n,k,q} = 1$ signifies that subcarrier \( n \) is allocated to user \( k \) at timeslot \( q \), while $\alpha_{n,k,q} = 0$ indicates that subcarrier \( n \) is not allocated to user \( k \) at timeslot \( q \).

\subsection{Problem Formulation}

\begin{subequations}\label{eq:2}
In this paper, our objective is to maximize the system sum rate by jointly optimizing the RB allocation decisions $\alpha_{n,k,q}$, beamforming vectors at the BS $\mathbf{w}_{n,q}$ and RIS reflection phase shift $\theta_{m,q}$, which is formulated as

	\begin{align}
		\max_{\theta_{m,q},\mathbf{w}_{n,q},\alpha_{n,k,q}} & \quad \sum_{k=1}^{K}R_{k}   \notag  \\
		\mathrm{s.t.} \enspace \quad \quad & \quad \alpha_{n,k,q}\in\{0,1\}, \quad \forall n,k,q,  \label{eq:1B}\\
		& \quad \sum_{k=1}^{K}\alpha_{n,k,q}\leq1, \quad \forall n,q,  \label{eq:1C}\\
		& \quad \sum_{n=1}^N\|\mathbf{w}_{n,q}\|^2\leq P_{\max}, \quad \forall q,  \label{eq:1D}\\
		& \quad \theta_{m,q}\in\{0,\pi\}, \quad \forall m,q,  \label{eq:1E}\\
		& \quad R_{k}\geq R_{\text{QoS}}, \quad \forall k, \label{eq:1F}
	\end{align}
\end{subequations}
 where \eqref{eq:1B} describes the binary constraints of the RB allocation decisions, \eqref{eq:1C} represents that each subcarrier can be allocated to at most one user. Constraint \eqref{eq:1D} governs the transmit power at the BS, ensuring it does not exceed the BS's available power $P_{\max}$. Constraint \eqref{eq:1E} is the discrete phase shift constraints on the RIS, where we assume the RIS is of 1-bit phase shift resolution in this paper, that is the phase shift of each reflection element can only be 0 or $\pi$. Constraint \eqref{eq:1F} addresses the system's QoS requirements.
 
 \par The optimization problem in \eqref{eq:2} is a non-convex combinatorial problem due to the non-convexity of the binary constraints in \eqref{eq:1B} and \eqref{eq:1E}, as well as the constraint \eqref{eq:1F}, which involves the rate $R_{k}$ of user $k$ under $\theta_{m,q}$. Therefore, traditional numerical optimization algorithms struggle to obtain high-quality solutions. To address this issue, in the following sections, we propose an UL-based joint resource allocation and beamforming design algorithm to effectively solve the optimization problem in \eqref{eq:2}.

\section{Unsupervised Learning Based Algorithm} \label{section:Unsupervised Learning Based Algorithm}
Recently, deep reinforcement learning (DRL) has gained significant popularity and been widely applied to numerous resource allocation tasks\cite{DRL, DRL1}. However, DRL relies on the Markov decision process (MDP), which inherently involves continuous interaction between the actions and the environment. This characteristic makes it unsuitable for the static optimization problem studied in this paper.
Therefore, in this section, we propose an UL-based approach to solve optimization problem in \eqref{eq:2}. Then, we present the details of the proposed approach.

\subsection{Overview of the Algorithm}

\begin{figure*}[!t]
	\centering
	
	\begin{minipage}[b]{\textwidth}
		\centering
		\subfloat[Proposed unsupervised learning algorithm network architecture]{\includegraphics[width=\textwidth]{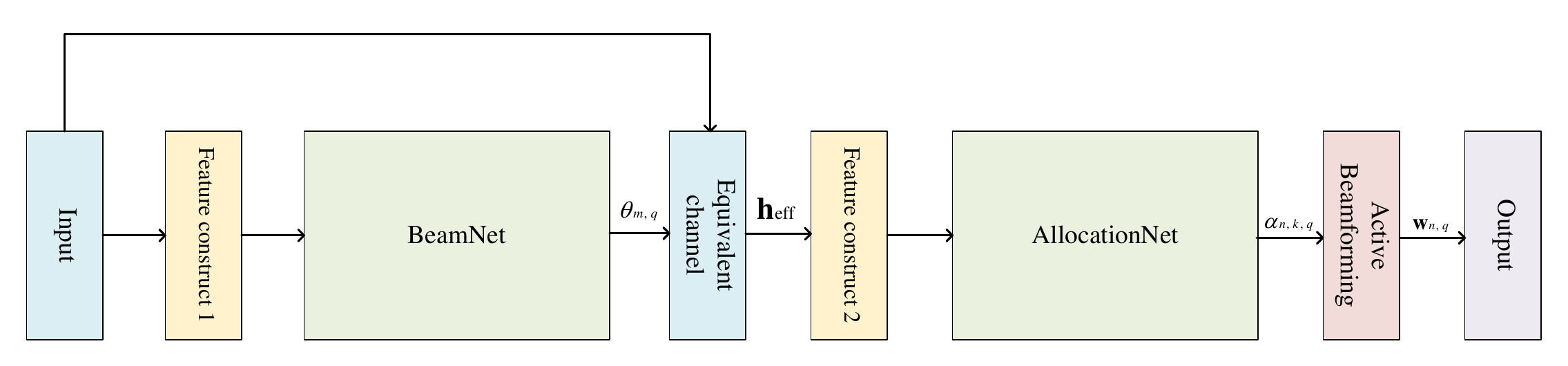} \label{Fig:total}}
		
	\end{minipage}
	\hspace{0.05\textwidth} 
	
	\begin{minipage}[b]{\textwidth}
		\centering
		\subfloat[BeamNet architecture]{\includegraphics[width=\textwidth]{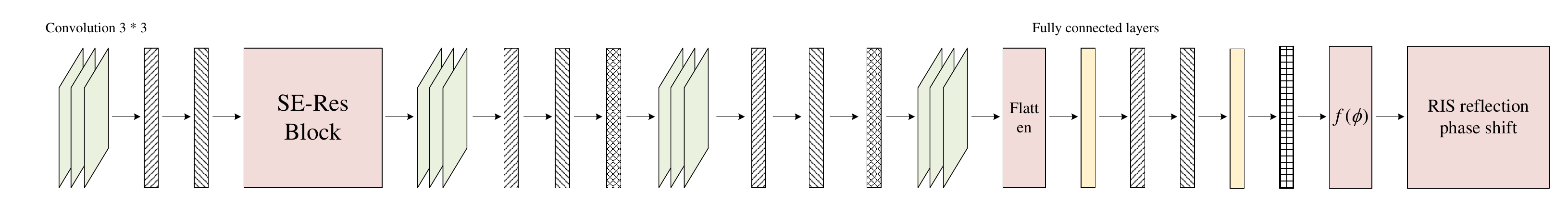} \label{BeamNet}}
		
	\end{minipage}
	\hspace{0.05\textwidth} 
	
	\begin{minipage}[b]{\textwidth}
		\centering
		\subfloat[AllocationNet architecture]{\includegraphics[width=\textwidth]{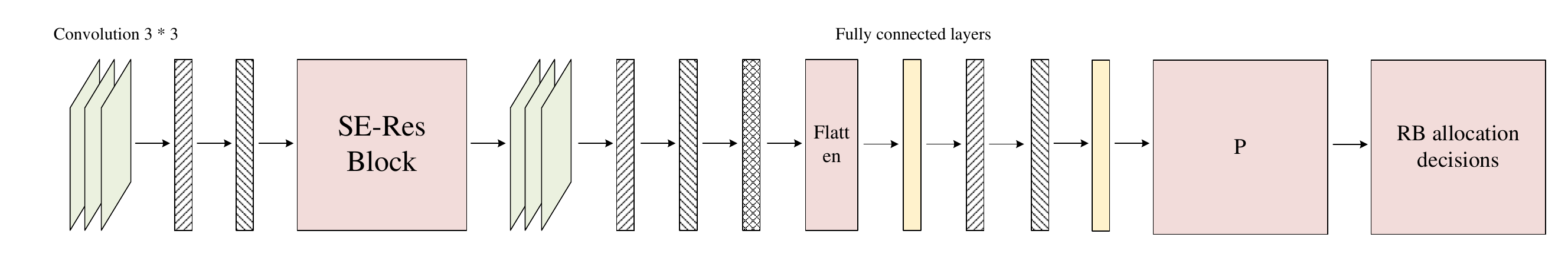} \label{AllocationNet}}
		
	\end{minipage}
	\hspace{0.05\textwidth} 
	
	\begin{minipage}[b]{\textwidth}
		\centering
		\subfloat[SE-Res block]{\includegraphics[width=0.7\textwidth]{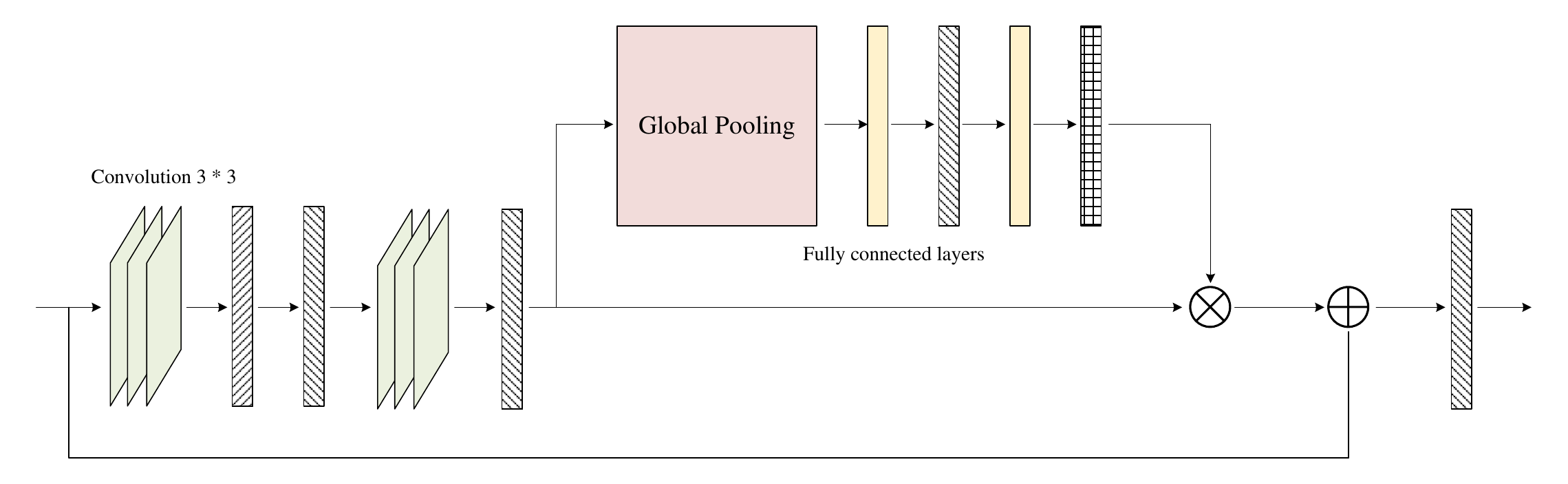}%
			\label{SE-Res}}
		\hfil
		\subfloat[Symbols]{\includegraphics[width=0.3\textwidth]{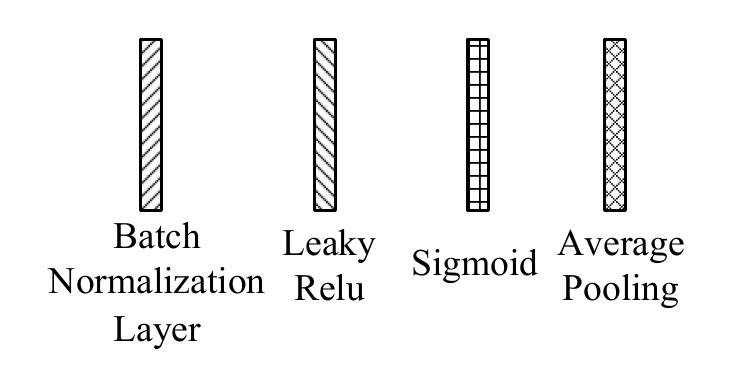}%
			\label{Symbols}}
		
	\end{minipage}
	
	\caption{Network architecture.}
	\label{fig:main}
\end{figure*}

In this section, a joint resource allocation and beamforming design algorithm based on UL is introduced. Since using a single neural network to output all variables would result in an excessive number of training parameters, this paper proposes a two-stage network architecture to maximize the system sum rate while satisfying the QoS constraints. This architecture is shown in Fig. \ref{fig:main}\subref{Fig:total}, it contains two neural networks, named as BeamNet and AllocationNet, where the BeamNet is utilized to predict the RIS reflection phase shift, while the AllocationNet is employed to output RB allocation decisions. In this paper, each subcarrier is allocated exclusively to one user, ensuring no mutual interference. Then, with the RIS reflection and RB allocation decisions produced by the neural network, the optimal active beamforming vectors $\mathbf{w}_{n,q}^*$ for the subcarrier $n$ at the $q$-th timeslot can be derived as maximum ratio transmission (MRT) in conjunction with the water-filling algorithm, i.e., 

\begin{equation}
	\mathbf{w}_{n,q}^*=\sqrt{p_{n,q}}\bar{\mathbf{w}}_{n,q}, \label{water}
\end{equation}
where $\bar{\mathbf{w}}_{n,q}$ represents the optimal unit-norm beamforming vector, which can be expressed as follows,
\begin{equation}
	\bar{\mathbf{w}}_{n,q}=\frac{\left(\bar{\mathbf{h}}^{(d)}_{n,k}+\boldsymbol{\varphi}_{\boldsymbol{q}}^T\bar{\mathbf{h}}^{(r)}_{n,k}\right)^H}{\|(\bar{\mathbf{h}}^{(d)}_{n,k}+\boldsymbol{\varphi}_{\boldsymbol{q}}^T\bar{\mathbf{h}}^{(r)}_{n,k})\|},\forall n,q ,
\end{equation}
and $p_{n,q}$ refers to the transmission power on subcarrier $n$ at timeslot $q$, which can be expressed by the following equation, 
\begin{equation}
	p_{n,q}=\left(\frac{1}{\tau_q}-\frac{1}{c_{n,q}}\right)^+, \forall n,q,
\end{equation} 
where $c_{n,q} = \frac{\|(\bar{\mathbf{h}}^{(d)}_{n,k}+\boldsymbol{\varphi}_{\boldsymbol{q}}^T\bar{\mathbf{h}}^{(r)}_{n,k})\|^2}{\sigma^2}$, and $\tau_q$ is the Lagrange multiplier that satisfies,
\begin{equation}
	\begin{aligned}
		\sum_{i=1}^N\left(\frac{1}{\tau_q}-\frac{1}{c_{i,q}}\right)=P_{\max}, \forall q. 
	\end{aligned}
\end{equation}

\subsection{BeamNet Design} \label{beam}

To predict the RIS reflection phase shift, BeamNet is proposed, as illustrated in Fig. \ref{fig:main}\subref{BeamNet}. The input to BeamNet is the channel state information (CSI), and the output is the RIS reflection phase shift matrix. Specifically, since the input data includes both the BS-User direct channel and the BS-RIS-User cascaded channel, its dimension is $N \times K \times N_t + N \times K \times M \times N_t$. Moreover, because neural networks cannot directly process complex numbers, the real and imaginary parts of the input data are separated. Using Feature construct 1, the data is reshaped into a four-dimensional real-valued tensor with dimensions $2K \times (M + 1) \times N \times N_t$, which is then fed into the neural network for further processing.

As shown in Fig. \ref{fig:main}\subref{BeamNet}, BeamNet consists of multiple convolutional layers, batch normalization layers, average pooling layers, fully connected layers, and an SE-Res block. The structure of the SE-Res block is depicted in Fig. \ref{fig:main}\subref{Symbols}, while the symbols representing the various layers are explained in Fig. \ref{fig:main}\subref{SE-Res}.

During forward propagation, the input tensor is first processed by convolutional layer, which extract channel features. The extracted features then pass through batch normalization layers that accelerate convergence and stabilize the training process. Next, average pooling layer reduce the dimensionality of the feature maps while preserving the most important information. These processed features are then fed into the SE-Res block, the core design of BeamNet, which significantly enhances the network’s representational capability.

The SE-Res block combines a residual structure with a squeeze-and-excitation mechanism: the residual connections mitigate the vanishing gradient problem, while the attention mechanism adaptively re-weights channel-wise features, enabling the network to focus on more informative paths. After the SE-Res block, two additional convolutional layers are applied to further refine and extract deeper features. Finally, the refined features are fed into fully connected layers to generate the predicted RIS reflection phase shift. 

To accommodate the hardware constraints of RIS, a quantization layer is introduced to discretize the RIS reflection phase shifts, which is essentially a step function. However, the step function is inherently non-differentiable, posing challenges for gradient computation and backpropagation during training. To overcome this, a differentiable approximation function is employed as an alternative to the non-differentiable step function. In particular, the 1-bit quantization function $f(\phi)$ is expressed as

\begin{figure}[t]
	\centering
	\includegraphics[width=0.5\textwidth, keepaspectratio]{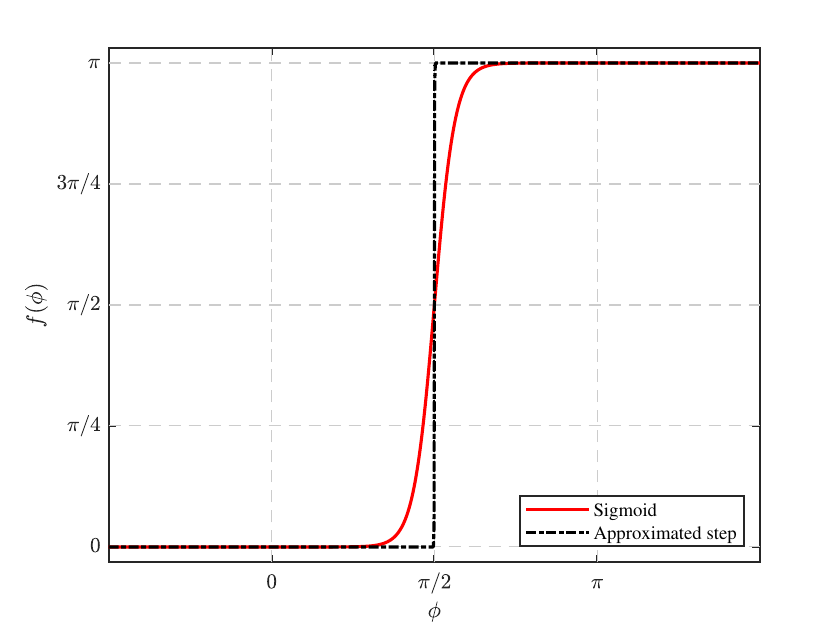}
	\caption{ The step function and the approximating function.}
	\label{Fig:step}
\end{figure}

\begin{equation}
	f\left(\phi\right)=\pi\mathrm{sigmoid}\left(\beta \left(\phi-\pi\right)\right),
\end{equation}
and the function image is shown in \Cref{Fig:step}, where \(\beta\) is a hyperparameter that determines the degree of deviation of the approximating function from the step function.

\subsection{AllocationNet Design} \label{allocation}
Based on the RIS reflection phase shift output by BeamNet, we can obtain the equivalent channel vector of BS-User link \(\mathbf{h}_{\rm eff}\), which can be expressed as

\begin{equation}
	\mathbf{h}_{\rm eff} = \bar{\mathbf{h}}^{(d)}_{n,k}+\boldsymbol{\varphi}_{{q}}^T\bar{\mathbf{h}}^{(r)}_{n,k},  \label{channel}
\end{equation}

\noindent where $\bar{\mathbf{h}}^{(d)}_{n,k}$ and $\bar{\mathbf{h}}^{(r)}_{n,k}$ represent the direct link channel and cascaded reflection channel, respectively, of user \( k \) on subcarrier \( n \) in the frequency domain.

The input to the AllocationNet is $\mathbf{h}_{\rm eff}$, which has dimensions \(N \times K \times N_t\), and the output is the subcarrier allocation decisions. Since neural networks cannot directly process complex numbers,  Feature construct 2 is utilized to reshape the $\mathbf{h}_{\rm eff}$ into a \(2 \times N \times K \times N_t\) dimensional data, which is then fed into the AllocationNet for further processing. 

The AllocationNet consists of multiple convolutional layers, batch normalization layers, average pooling layers, fully connected layers, and an SE-Res block, with its overall structure depicted in Fig. \ref{fig:main}\subref{AllocationNet}. The SE-Res block and symbols are shown separately in Fig. \ref{fig:main}\subref{SE-Res} and Fig. \ref{fig:main}\subref{Symbols}, respectively.

The AllocationNet uses the Gumbel softmax trick and ensures the ability to back propagate while maintaining discrete outputs and effectively satisfies constraints \eqref{eq:1B} and \eqref{eq:1C} in \eqref{eq:2}. Specifically, Gumbel softmax trick is a reparameterization technique that generates outputs close to one-hot vectors by adding Gumbel noise to the logits (unnormalized probabilities) and processing them with the softmax function. 

The output of the fully connected layer FC2 is reconstructed to produce a probability matrix \( \mathbf{P} \in \mathbb{R}^{N \times K  \times Q} \), where each slice of \( \mathbf{P} \) along the second dimension (\( K \)) represents the probabilities of allocating \( N \) subcarriers across \( Q \) timeslots to user \( k \). To intuitively select subcarriers, in each time slot, each subcarrier is allocated to the user with the highest probability by applying the \(\text{argmax}\) function, resulting in a one-hot encoded selection tensor. However, as the \(\text{argmax}\) function is non-differentiable and does not support backpropagation, the Gumbel softmax trick is introduced as a differentiable approximation, enabling gradient-based optimization during training. The detailed formulation is as follows\cite{jang2016categorical}

\begin{equation}
	\alpha_{n,k,q}=\frac{\exp\big((\mathbf{P}(n,k,q)+g_{n,k,q})/\tau\big)}{\sum\limits_{k'=1}^{K}\exp\left((\mathbf{P}(n,k',q)+g_{n, k',q})/\tau\right)},
\end{equation}
where \( \tau \) denotes the temperature parameter. As \( \tau \) approaches zero, the Gumbel softmax output increasingly resembles a one-hot vector. However, excessively small values of \( \tau \) can result in the gradient vanishing issues, which necessitates careful tuning of \( \tau \) to achieve an optimal trade-off between model performance and training efficiency. In addition,  \( g_n \) refers to Gumbel noise, which is characterized by the following probability distribution

\begin{equation}
	g_n = - \ln(-\ln(u)),
\end{equation}

\noindent where, \(u\sim\mathbf{U}(0,1)\) follows a uniform distribution. Finally, the RB allocation decisions can be obtained from the output of the AllocationNet network.

\subsection{Customized Loss Function}
The proposed customized loss function is comprised of two distinct components: the optimization objective and the regularizers. In this paper, the objective is to maximize the system sum rate. Consequently, this optimization objective component is defined as
\begin{equation}
	\mathcal{L}_{\mathrm{rate}} = -  \sum_{k=1}^K R_k
	= - \frac{W}{Q} \sum_{k=1}^K \sum_{q=1}^Q \sum_{n=1}^N \alpha_{n,k,q} \log_2(1+\gamma_{n,k,q}).
\end{equation}

\par 
Given that Equation \eqref{eq:1F} is an inequality constraint related to QoS, we introduce a penalty term to ensure that the final output satisfies this constraint. The role of the penalty term is to increase the objective function's penalty value when the solution does not meet the QoS constraint. This compels the optimization process to adjust the distribution of solutions, gradually satisfying the QoS requirements. Specifically, the penalty term serves to penalize non-compliant solutions in the objective function, thereby guiding the optimization algorithm to prioritize solutions that fulfill the QoS constraint. Through this mechanism, the penalty term not only enhances the robustness of the model but also ensures the practical feasibility and effectiveness of the optimization results\cite{TowardsLiang}.  The penalty term is given by
$
	\mathcal{L}_{\text{QoS}} = \lambda_1 \sum_{k=1}^{K} (R_{\text{QoS}}-R_{k},0)^{+},
$
where $\lambda_1$ is a hyperparameter that can be adjusted according to different requirements.
Furthermore, to prevent overfitting, we impose regularization on the network parameters, i.e., 
$
	\mathcal{L}_{\mathrm{2}} =\lambda_2\|\Theta_{\mathrm{Network}}\|^2,
$
where $\Theta_{\mathrm{Network}}$ denotes all trainable network parameters, $\lambda_2$ is the weight factor. Typically, $\lambda_2$ is set to a small value to balance the model's complexity and generalization ability.

\par
Therefore, the loss function is then defined as follows
\begin{equation}
	\mathcal{L} = \mathcal{L}_{\mathrm{rate}} + \mathcal{L}_{\text{QoS}} + \mathcal{L}_{\mathrm{2}}.
\end{equation}

In practical applications, neural networks often employ mini-batch update strategies to balance computational efficiency with convergence stability during training. Therefore, the mini-batches loss function is defined as follows
\begin{equation}
	Loss =\frac{1}{B}\sum_{b=1}^{B}\mathcal{L}^{(b)},  \label{loss}
\end{equation}
where $B$ is the size of samples in a mini-batch, $\mathcal{L}^{(b)}$ represents the loss function value of the $b$-th sample.

\begin{algorithm}[!t]
	\caption{UL-based joint resource allocation and beamforming design algorithm}
	\label{alg:1}
	\begin{algorithmic}[1]
		\STATE \textbf{Input:} $\mathbf{\bar{h}}^{(d)}_{n,k}$ and $\mathbf{\bar{h}}^{(r)}_{n,k}$.
		\STATE \textbf{Output:} $\theta_{m,q}$, $\alpha_{n,k,q}$ and $\mathbf{w}_{n,q}$.
		\STATE Initialize network parameters.
		\FOR{$i = 1, 2, 3, \ldots, N_5$}
		\STATE Randomly sample mini batches $B$ data from training dataset.
		\STATE Features construction according to \cref{beam}.
		\STATE Enter features data into BeamNet to get $\theta_{m,q}$.
		\STATE Compute the equivalent channel $\mathbf{h}_{\rm eff}$ using \eqref{channel}.
		\STATE Features construction according to \cref{allocation}.
		\STATE Enter features data into AllocationNet to get $\alpha_{n,k,q}$.
		\STATE According to $\theta_{m,q}$, $\alpha_{n,k,q}$, and $\mathbf{h}_{\rm eff}$, calculate $\mathbf{w}_{n,q}$ using \eqref{water}.
			\STATE Compute the loss function \eqref{loss}.
		
		\IF{$1 \leq i \leq N_1$}
			\STATE Update the BeamNet.
		\ELSIF{$N_1 < i \leq N_2$}
			\STATE Update the AllocationNet.
		\ELSIF{$N_2 < i \leq N_3$}
			\STATE Update the BeamNet.
		\ELSIF{$N_3 < i \leq N_4$}
			\STATE Update the AllocationNet.
		\ELSE
			\STATE Jointly update the BeamNet and AllocationNet.
		\ENDIF
		\ENDFOR
	\end{algorithmic}
\end{algorithm}

\subsection{Network Training} \label{train}

Since the BeamNet and AllocationNet share the same loss function, it is possible to jointly train the entire network.  However, due to the large number of parameters in the entire net, direct joint training from scratch is inefficient.  Therefore, this paper proposes a phased training approach to optimize the entire network.

The entire training process adopts the Adam optimizer across all stages. Adam is an adaptive learning rate optimizer that adjusts the learning rate for each parameter individually based on estimates of first and second moments of the gradients. The training is divided into five stages, each using a different setting of the initial learning rate. In the first stage, BeamNet is trained with an initial learning rate of $\mu_1$; in the second stage, AllocationNet is trained with $\mu_2$. The third and fourth stages involve further training of BeamNet and AllocationNet, respectively, using the same initial learning rates as before. During the individual training of one sub-network, the parameters of the other remain fixed. In the fifth and final stage, the entire network is jointly trained with a smaller initial learning rate $\mu_3$. The complete training procedure is summarized in \Cref{alg:1}. Typically, we set $\mu_3 < \mu_1$ and $\mu_3 < \mu_2$; for simplicity, we use $\mu_3 = 0.5 \mu_1 = 0.5 \mu_2$ in this work. This approach significantly improves training efficiency and accelerates convergence toward a high-quality solution.

\section{Numerical Results}   \label{section:Numerical Results}

In this section, we thoroughly evaluate the performance of the proposed algorithm. To assess its effectiveness, we conduct a series of extensive simulations and compare the results with those obtained from several existing methods. The evaluation focuses on key performance metrics, including sum rate, robustness, and complexity, to provide a comprehensive insight into the algorithm's potential.

\subsection{Simulation Settings}
\begin{figure}[t]
	\centering
	\includegraphics[width=0.5\textwidth, keepaspectratio]{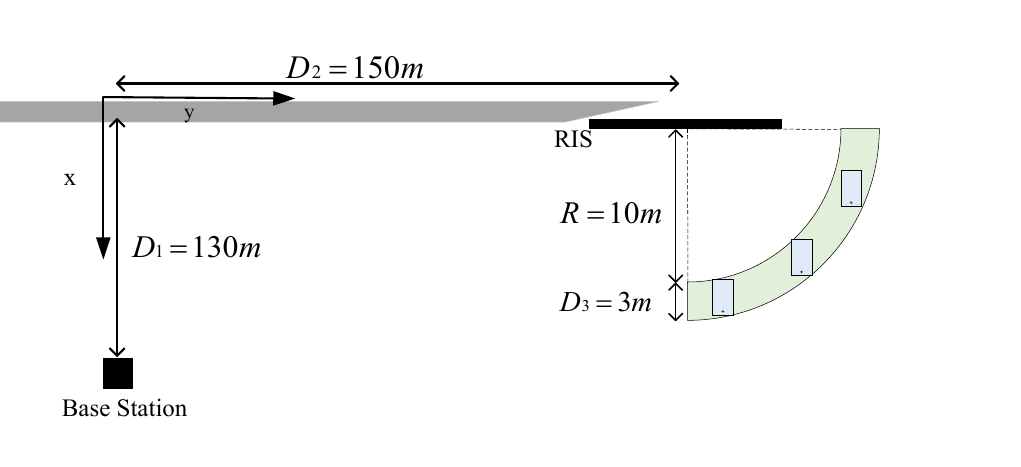}
	\caption{Illustration of the simulated scenario. }
	\label{Fig:position}
\end{figure}

As illustrated in \Cref{Fig:position}, a downlink MISO-OFDMA system with three users is considered, where the users are located in a quarter annular region with inner and outer radii of 10 m and 13 m, respectively. In the x-axis direction, the distance between the BS and the origin is \(D_1 = 130\) m, and in the y-axis direction, the distance between the origin and the RIS is \(D_2 = 150\) m. The number of reflection elements on the RIS is \(M=\) 64, the number of timeslots in the dynamic allocation scheme is \(Q=\) 6, the number of subcarriers is \(N = \) 16, and the BS is equipped with 4 antennas. Furthermore, a noise power spectral density of \(-174 \ \text{dBm/Hz}\) and each subcarrier bandwidth of \(180 \ \text{kHz}\) are assumed. The QoS rate is configured at 2 Mbps. The distance between two adjacent antennas on the BS and the distance between adjacent reflection elements on the RIS are all set to be half the carrier wavelength. The delay taps are set to $L_0 = 4$, $L_1 = 2$, and $L_2 = 3$, respectively. Additionally, for the direct BS-user channel, Rayleigh fading channels are assumed, while Rician fading is considered for the reflection channels from BS to RIS and from RIS to users. The first tap of the reflection channel is set to be the line-of-sight (LoS) path, and the remaining taps are non-line-of-sight (NLoS) paths. The Rician factors for the BS-RIS link and RIS-User link are respectively represented by

\begin{table}
	\begin{center}
		\caption{Hyperparameter Description and Values.}
		\label{tab1}
		
		\begin{tabular}{| c | c | c |}
			\hline
			Parameter & Description & Value\\
			\hline
			$\mu_1$ & Learning rate for training BeamNet update & 0.001\\
			\hline
			$\mu_2$ & Learning rate for training AllocationNet update & 0.001\\ 
			\hline
			$\mu_3$ & Learning rate for training JointNet update & 0.0005\\ 
			\hline
			$\lambda_1$ &  Softmax temperature for gumbel softmax & 5\\ 
			\hline
			$\lambda_2$ &  Regularization Parameter & 5e-5\\ 
			\hline
			$\tau$ &  Temperature for gumbel softmax & 0.5\\ 
			\hline
			& The degree of difference between the sum of    &   \\
			$\beta$ &   shifted sigmoid functions and the step  & 100\\
			&    quantization function &   \\
			\hline 
			$D$ &  mini-batch size & 32\\ 
			\hline
		\end{tabular}
		
	\end{center}
\end{table}

\begin{equation}
	{k}_{\text{BR}}=\frac{P_{\text{LoS,BR}}}{P_{\text{NLoS,BR}}},\quad{k}_{\text{RU}}=\frac{P_{\text{LoS,RU}}}{P_{\text{NLoS,RU}}},
\end{equation}

\noindent where, \(P_{\text{LoS, BR}}\), \(P_{\text{NLoS, BR}}\), \(P_{\text{LoS, RU}}\), and \(P_{\text{NLoS, RU}}\) represent the power of the LoS and NLoS paths for the corresponding links. In this simulation, \({k}_{\text{BR}}=2\) dB and \({k}_{\text{RU}}=4\) dB are assumed. The large-scale fading is defined as \( \beta = \beta_0 \left(\frac{d}{d_0}\right)^{-\xi} \), where \( \beta_0 = -30 \) dB, \( d_0 = 1 \) m, \( d \) represents the link distance, and \( \xi \) is the path loss exponent. The path loss exponents for the BS-User link, BS-RIS link, and RIS-User link are \( \xi_0 = 3.8 \), \( \xi_1 = 2.2 \), and \( \xi_2 = 2.4 \), respectively.

For the training and evaluation of the network, 4,900 and 100 data samples were generated as the training and validation sets, respectively. The result are averaged over 100 channel realization. The specific network architecture is shown in Fig. \ref{fig:main}\subref{Fig:total}, where \(N_1= 2500\), \(N_2 = 5000\), \( N_3 = 7000\), \(N_4 = 9000\), and \(N_5 = 15000\). The remaining hyperparameters are listed in \Cref{tab1}.

We compare the following methods in the simulation:

\begin{itemize}
	\item \textbf{Continuous SCA}: The algorithm proposed in \cite{OFDMA_zhangRui} is employed for RB allocation and RIS reflection phase shift optimization, while MRT is used for active beamforming at the BS.
	\item \textbf{Discrete SCA}: In this scheme, the RIS reflection phase shift from continuous SCA are directly quantized to achieve discrete phase settings.
	\item \textbf{Proposed continuous algorithm}: The proposed \Cref{alg:1} is used, but the quantization layer is not implemented in BeamNet, resulting in output values from BeamNet that consist of continuous values ranging from 0 to \(2\pi\).
	\item \textbf{Proposed discrete algorithm}: The proposed \Cref{alg:1}.
	
	\item \textbf{Random allocation}: The RB allocation decisions are randomly set, while active beamforming and RIS reflection phase shift are optimized using the proposed algorithm.
	
	\item \textbf{Random RIS}: The RIS reflection phase shift are set randomly, while RB allocation decisions and BS beamforming are optimized using the proposed algorithm. Specifically, the RIS reflection phase shift are randomly selected between 0 and $\pi$.
	\item \textbf{Without RIS}: This scheme represents a system that lacks RIS assistance, i.e., the number of RIS reflection elements is set to \( M = 0 \).

\end{itemize}

\subsection{Sum Rate versus Transmit Power}

\begin{figure}[t]
	\centering
	\includegraphics[width=0.5\textwidth, keepaspectratio]{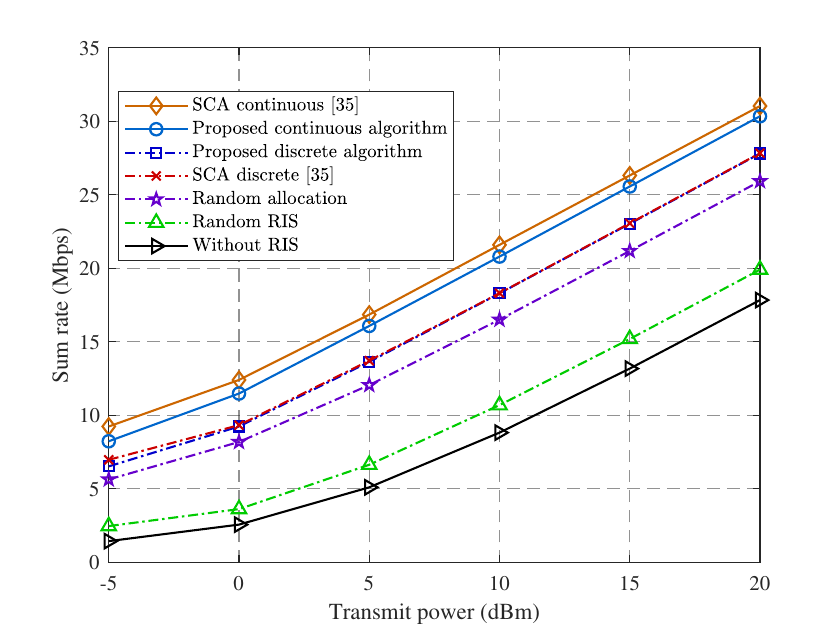}
	\caption{Sum rate versus transmit power.}
	\label{Fig:power}
\end{figure}

\Cref{Fig:power} compares the sum rates achieved by different algorithms under varying transmit powers. It can be observed that all schemes employing RIS significantly outperform the without RIS benchmark, confirming the capability of RIS to enhance the performance of wireless communication systems. This improvement is mainly attributed to the passive beamforming gain introduced by RIS, as evidenced by the Random RIS benchmark, which already provides a notable performance gain over the without RIS case \cite{gain}. Furthermore, the additional improvement achieved by the optimized RIS schemes over the random RIS scheme further validates the importance of intelligent phase control. The proposed continuous algorithm achieves a sum rate close to that of the continuous SCA benchmark, demonstrating its near-optimal performance. Under discrete phase shift constraints, the proposed discrete algorithm also performs extremely close to the discrete SCA algorithm, indicating its effectiveness in handling discrete RIS phase shift control. Moreover, the proposed discrete algorithm outperforms the Random allocation scheme, which highlights the critical role of RB allocation in RIS-assisted systems. In general, continuous phase shifts offer finer RIS reflection phase shift control and beamforming control, thereby providing performance advantages over discrete phase shifts. However, due to hardware cost and implementation complexity, practical RIS deployments typically adopt low-resolution discrete phase shifters rather than ideal continuous ones, making efficient algorithm design under quantization constraints essential\cite{dis}.

\begin{figure*}[!t]
	\centering
	\subfloat[Different penalty factors versus 5\% rate.]{\includegraphics[width=0.5\textwidth]{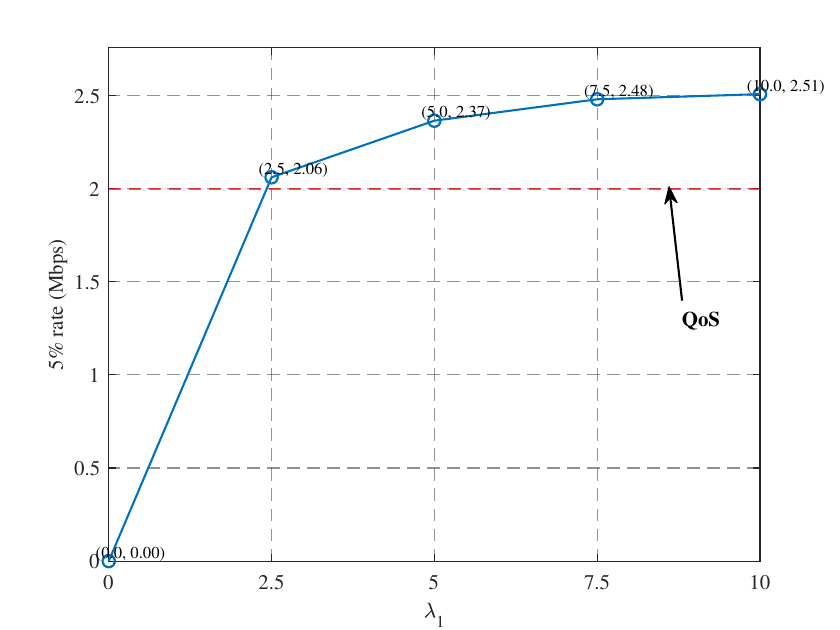}%
		\label{lamdba1}}
	\hfil
	\subfloat[ Different penalty factors versus system sum rate.]{\includegraphics[width=0.5\textwidth]{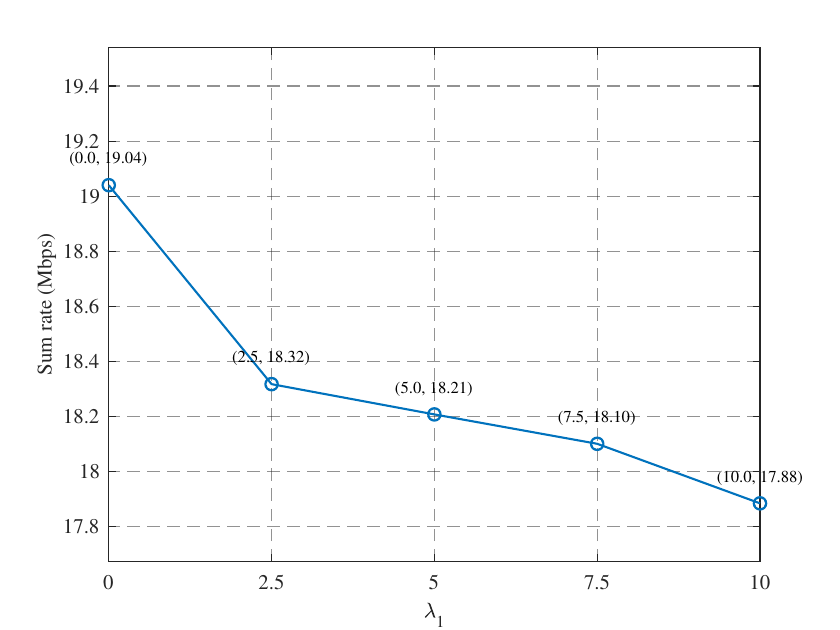}%
		\label{lamdba2}}
	\caption{The impact of different penalty factors on the proposed algorithm.}
	\label{lamdba}
\end{figure*}

\begin{figure*}[!t]
	\centering
	\subfloat[Loss value versus iteration under different learning rates.]{\includegraphics[width=0.5\textwidth]{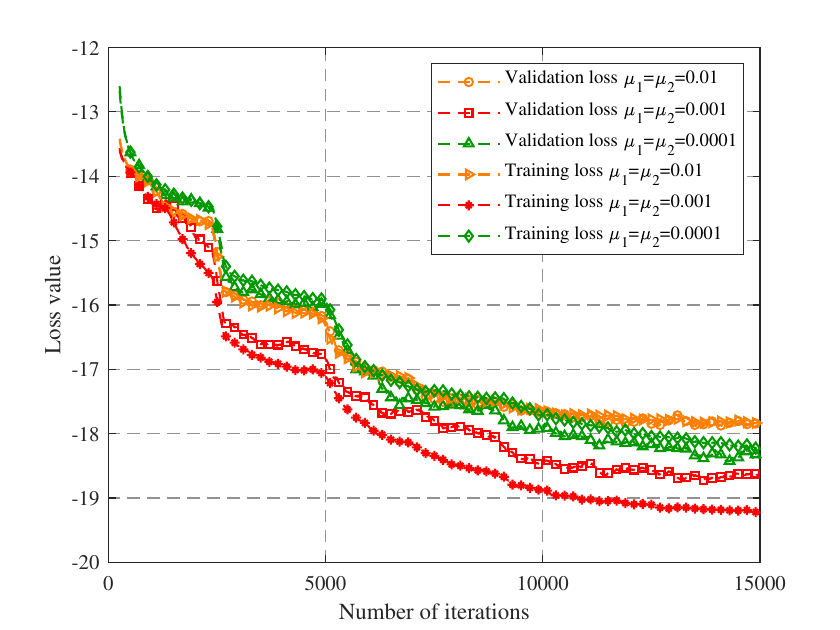}%
		\label{Fig:learningrate}}
	\hfil
	\subfloat[Loss value versus different training methods.]{\includegraphics[width=0.5\textwidth]{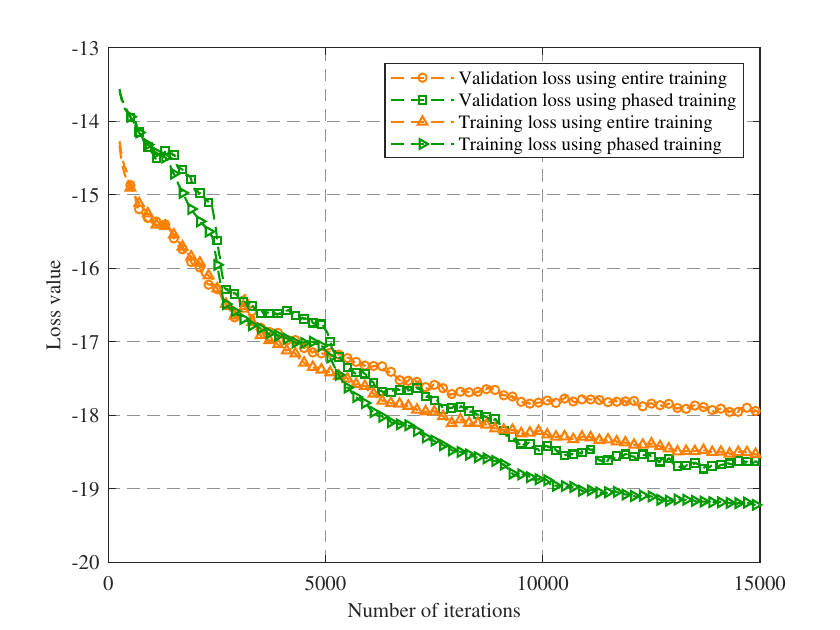}%
		\label{Fig:learningratevs}}
	\caption{The impact of different learning rates and training methods on the proposed algorithm.}
	\label{learningrate}
\end{figure*}

\subsection{Impact of Different Penalty Factor on Proposed Algorithm}

\Cref{lamdba} illustrates the impact of different penalty factors \(\lambda_1\) on the performance of the proposed algorithm. To evaluate the effect of \(\lambda_1\) on the system sum rate and the satisfaction of the QoS constraint, we compares the 5th percentile rate of the system under different $\lambda_1$, which refers to  the users' in the bottom 5\% when the rate of all users rates are sorted from high to low \cite{LearningNaderiAlizadeh}. In Fig. \ref{lamdba}\subref{lamdba1}, as the penalty factor \(\lambda_1\) increases, the 5th percentile rate significantly improves, surpassing the QoS threshold. However, when $\lambda_1$ reaches higher values, the rate increase slows down, indicating that a moderate $\lambda_1$ is beneficial for enhancing the performance of the worst-case users in the system. An excessively large $\lambda_1$ distorts the training objective, causing the model to overly focus on the penalty term while neglecting the original optimization goal. Notably, when \(\lambda_1 = 5\), the 5th percentile rate exceeds the QoS constraint, indicating that nearly all users meet the QoS constraint.  Fig. \ref{lamdba}\subref{lamdba2} shows the variation in the system sum rate as \(\lambda_1\) changes. As \(\lambda_1\) increases, the sum rate gradually decreases. Combining the observations from  Fig. \ref{lamdba}\subref{lamdba1} and  Fig. \ref{lamdba}\subref{lamdba2} suggests that while larger \(\lambda_1\) values can effectively enhance the 5th percentile rate, it negatively affects the overall system performance. This can be attribute to the fact that as the penalty factor increases, the network becomes more focused on the penalty term in the optimization process, leading to a reduction in the system's sum rate. Therefore, the selection of the penalty factor \(\lambda_1\) requires a balance between improving the performance of the worst-case users and maintaining overall system performance. Considering both the QoS requirements and the system sum rate, we use in the followings simulations \(\lambda_1 = 5\) as the optimal penalty factor.

\subsection{Impact of Learning Rate on Proposed Algorithm}

Fig. \ref{learningrate}\subref{Fig:learningrate} illustrates the impact of different initial learning rates $\mu_1$, $\mu_2$ (0.01, 0.001, 0.0001) on training and validation losses over the course of iterations. The results indicate that when the initial learning rate is set to 0.01, both the training and validation losses decrease rapidly in the initial stages. However, as the number of iterations increases, the decline in loss gradually slows down, and the model's loss ultimately converges to the highest value. In contrast, an initial learning rate of 0.001 achieves a good balance between the speed of loss reduction and stability, with both training and validation loss curves appearing smooth and ultimately converging to the lowest values. This demonstrates that an initial learning rate of 0.001 effectively optimizes the model while maintaining good generalization performance. With an initial learning rate of 0.0001, although the training and validation loss curves exhibit the most stable trends, the convergence speed is slow, and the final loss value is higher compared to the case with 0.001, suggesting that this lower learning rate may lead to underfitting. Therefore, an initial learning rate of 0.001 performs best in this experiment, ensuring optimization stability while achieving lower loss values.

Fig. \ref{learningrate}\subref{Fig:learningratevs} illustrates a performance comparison between different training methods when the initial learning rate is set to 0.001. The network using the phased training method was trained following the procedure outlined in \cref{train}, while the entire training method involves training the BeamNet and AllocationNet jointly as a unified system. Although the network trained using the phased training method shows a slower decline in loss during the early stages, it is able to converge to a lower and more stable value compared to the network trained with the entire training method. These results highlight the effectiveness of the phased approach in optimizing network performance over time. This is mainly because the joint network has a large number of parameters, which affects the convergence speed. In addition, the two networks are responsible for very different tasks, thus requiring different learning policies. The phased training strategy accommodates these differences and allows each network to focus on its own optimization objective, leading to improved overall performance.

\subsection{Impact of Reflection Elements $M$}

\begin{figure}[t]
	\centering
	\includegraphics[width=0.5\textwidth, keepaspectratio]{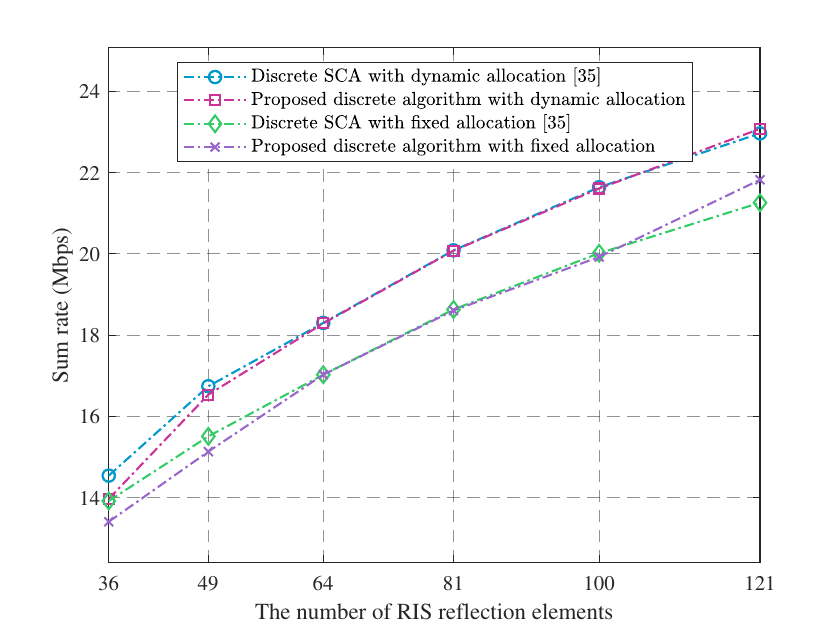}
	\caption{The impact of the number of reflection elements on the system sum rate, with training and testing conducted under BS power constraint of $P_{\max} = 10$ dBm.}
	\label{Fig:element}
\end{figure}

\begin{figure}[t]
	\centering
	\includegraphics[width=0.5\textwidth, keepaspectratio]{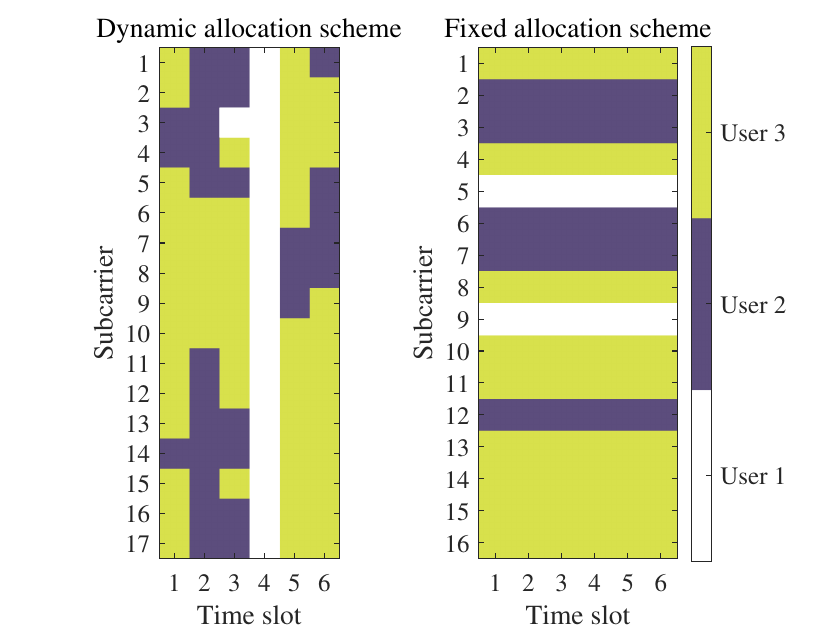}
	\caption{Comparison of various RB allocation schemes: the RB colored in white/purple/yellow represents that it is allocated to User 1/2/3. }
	\label{Fig:allocation}
\end{figure}

\Cref{Fig:element} illustrates the impact of the number of RIS elements $M$ on the system sum rate. As $M$ increases, the performance of all algorithms improves, and the proposed algorithm closely matches the performance of the SCA method, demonstrating its effectiveness. Both the dynamic allocation scheme and the fixed allocation scheme benefit from the increase in $M$. This is because a larger number of RIS elements provides greater beamforming flexibility and higher passive beamforming gain.
The observed difference in performance between the two allocation schemes as $M$ increases can be better understood with the aid of \Cref{Fig:allocation}, which shows an example of the RB allocation pattern under the proposed dynamic allocation and fixed allocation schemes. For the dynamic allocation scheme, RBs are assigned to fewer users in each timeslot, and in some cases, to only a single user. This behavior stems from the limitations of RIS phase shift control, which constrain its ability to simultaneously accommodate the channel conditions of multiple users. By serving fewer users per timeslot, higher beamforming gain can be obtained as the phase shift are customized for fewer channels, thereby maximizing beamforming gain. As the number of RIS elements increases, this gain becomes more significant, further widening the performance gap between the dynamic and fixed allocation schemes.

\subsection{Robustness Validation}

\begin{figure}[t]
	\centering
	\includegraphics[width=0.5\textwidth, keepaspectratio]{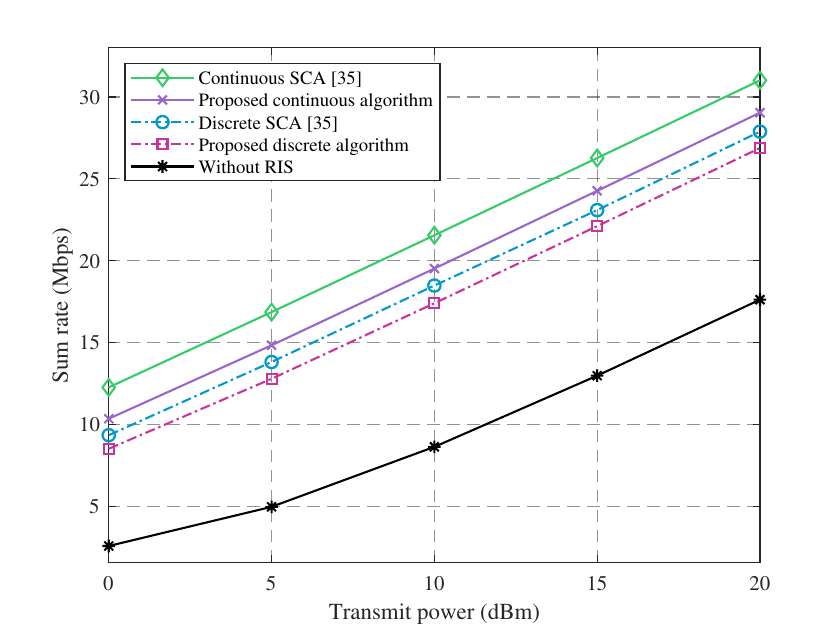}
	\caption{The sum rate versus transmit power evaluated with $L_0 = 6$, $L_1 = 3$, and $L_2 = 4$ and trained with $L_0 = 4$, $L_1 = 2$, and $L_2 = 3$.}
	\label{Fig:generalize_muti_path}
\end{figure}

\begin{figure}[t]
	\centering
	\includegraphics[width=0.5\textwidth, keepaspectratio]{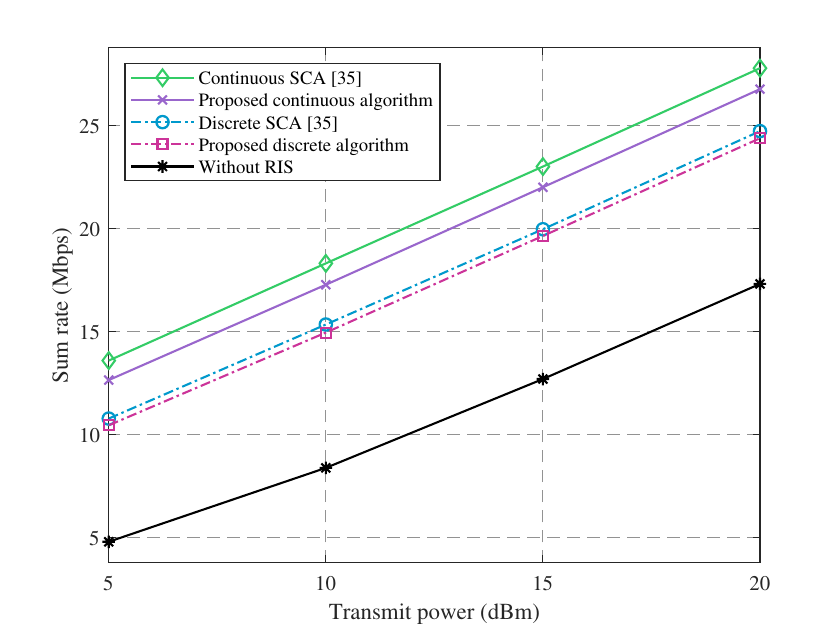}
	\caption{Sum rate versus transmit power, tested at \( R = 15 \, \text{m} \), \( D_3 = 3 \, \text{m} \), and trained at \( R = 10 \, \text{m} \), \( D_3 = 3 \, \text{m} \).}
	\label{Fig:generalize_position}
\end{figure}

\begin{figure}[t]
	\centering
	\includegraphics[width=0.5\textwidth, keepaspectratio]{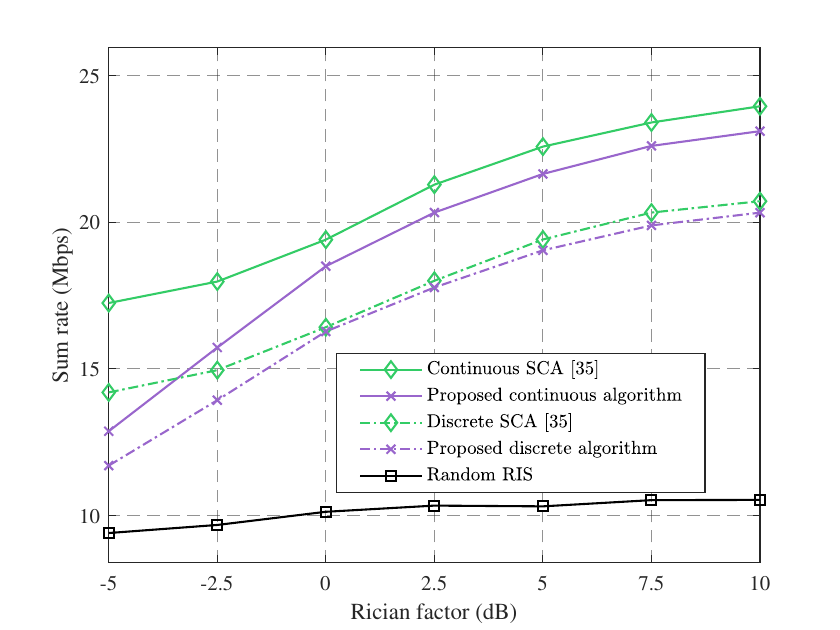}
	\caption{Sum rate versus Rician factor for different resolutions, with training at \({k}_{\text{BR}} = 2 \enspace \text{dB}\) and \({k}_{\text{RU}} = 4 \enspace \text{dB}\).}
	\label{Fig:RiceFactor}
\end{figure}

In this subsection, we demonstrate the robustness of the proposed method under different delay tap numbers, user distributions, and Rician factor in \figrange{Fig:generalize_muti_path}{Fig:RiceFactor}.

As shown in \Cref{Fig:generalize_muti_path}, the delay tap numbers in the testing environment are set to $L_0 = 6$, $L_1 = 3$, and $L_2 = 4$, while the network is trained using datasets with delay tap numbers $L_0 = 4$, $L_1 = 2$, and $L_2 = 3$. 
The results demonstrate that the proposed continuous and discrete algorithms generalize well to unseen channel conditions with different delay tap settings. 
Specifically, under a transmit power of 10 dBm, the proposed continuous algorithm achieves 90.55\% of the performance of the continuous SCA benchmark, while the proposed discrete algorithm attains 94.7\% of the performance of the discrete SCA benchmark. These results indicate that the proposed algorithms closely approximate their respective SCA-based counterparts.
Furthermore, the continuous schemes consistently outperform the discrete ones across all cases, highlighting the performance gain from continuous optimization. In addition, all RIS-assisted schemes significantly outperform the baseline without RIS, confirming the effectiveness of RIS in enhancing system performance.

In \Cref{Fig:generalize_position}, the user distributions are set to \( R = 15\) m and \( D_3 = 3 \) m, while the network is trained on datasets where the user distributions are set to \( R = 10 \) m and \( D_3 = 3 \) m. The results demonstrate that the trained network also performs well when \( R\) is different from the training dataset. Specifically, under the transmit power of 10 dBm, the proposed continuous and discrete algorithms achieve 94.33\% and 97.44\% of the performance of the continuous SCA and discrete SCA, respectively. This confirms the robustness of the proposed method under varying user position conditions.

\Cref{Fig:RiceFactor} illustrates the trend of system sum rate performance as the Rician factor varies. To simplify the analysis, we set ${k}_{\text{BR}} = {k}_{\text{RU}}$. During the training phase, the network was trained under conditions where the Rician factors are ${k}_{\text{BR}} = 2 \enspace \text{dB}$ and ${k}_{\text{RU}} = 4 \enspace \text{dB}$, with the BS transmit power $P_{\max}$ set to 10~dBm.
It can be observed that as the Rician factor increases, the performance of all algorithms consistently improves. 

This is because a higher Rician factor corresponds to greater channel 
sparsity. Under such conditions, the RIS reflection phase shifts can more easily align with the dominant signal path, thereby providing higher passive beamforming gain. Consequently, the received signal power increases, enhancing the SNR and improving the overall system sum rate.
When the Rician factor exceeds 0 dB, the proposed continuous and discrete algorithms demonstrate strong generalization ability, with only minor performance gaps compared to the continuous and discrete SCA algorithms. However, when the Rician factor is below 0 dB, the performance gap becomes more noticeable. Specifically, at $-2.5$ dB, the proposed discrete algorithm performs approximately 1.5 Mbps worse than the discrete SCA, while the continuous counterpart shows a gap of about 2.2 Mbps relative to the continuous SCA. Nevertheless, both still significantly outperform the RIS random scheme.
This performance degradation in low-Rician environments can be attributed to the mismatch between training and testing conditions. The network was trained under relatively strong LoS conditions, making it less effective at feature extraction in rich-scattering environments. In such cases, more resources may be allocated to users with poor channel conditions to satisfy QoS constraints, thereby limiting the system-wide performance.
However, in most practical RIS deployment scenarios, the communication links are typically dominated by LoS components, corresponding to higher Rician factors. Thus, the proposed algorithm remains highly robust and applicable in real-world settings.

\subsection{Complexity Analysis}

To compare the computational complexity of different algorithms, both training and testing were conducted on a platform equipped with an NVIDIA RTX 3060 GPU and an Intel i7-12700 CPU. The computational complexity of BeamNet is given by $\mathcal{O}(K^2 N_t N M + Q M)$, while that of AllocationNet is $\mathcal{O}(Q^2 N_t N K + Q N K)$. The complexity of the active beamforming algorithm is $\mathcal{O}(Q N M N_t + Q N)$. The computational complexity of the conventional SCA algorithm can be found in \cite{OFDMA_zhangRui}.
\Cref{tab2} reports the runtime and performance of each algorithm. In both continuous and discrete phase scenarios, the proposed algorithm significantly outperforms the SCA algorithm in terms of runtime while maintaining competitive performance. As $M$ increases, the computational cost of the SCA-based methods rises sharply due to their iterative nature. In contrast, the runtime of the proposed algorithms grows much more slowly.
Specifically, when $M = 64$, the proposed discrete algorithm achieves 99.93\% of the performance of the Discrete SCA algorithm, while requiring only 0.036\% of its runtime. This result highlights the efficiency of the proposed method. The substantial reduction in computational time is mainly attributed to the fact that our algorithm only requires a single forward pass through the neural network to produce the solution, without relying on time-consuming iterative optimization procedures.

\section{Conclusion}  \label{section:Conclusion}

\begin{table}[t]
	\begin{center}
		\caption{Performance and Runtime Comparison.}
		\label{tab2}
		
		\begin{tabular}{|c|c|c|c|}
			\hline
			$M$ & Algorithm & Runtime (ms) & Performance\tablefootnote{The benchmark performance for the system sum rate is set by the discrete SCA algorithm.} \\
			\hline
			\multirow{4}{*}{36} 
			& Continuous SCA & 19653.46 & 118.05\% \\
			& Discrete SCA & 20357.91 & 100\% \\
			& Proposed continuous & 17.69 & 112.98\% \\
			& Proposed discrete & 17.58 & 96.08\% \\
			\hline
			\multirow{4}{*}{64} 
			& Continuous SCA & 47356.23 & 117.96\% \\
			& Discrete SCA & 49642.81 & 100\% \\
			& Proposed continuous & 18.35 & 113.60\% \\
			& Proposed discrete & 18.21 & 99.93\% \\
			\hline
			\multirow{4}{*}{100} 
			& Continuous SCA & 114356.24 & 116.68\% \\
			& Discrete SCA & 118533.43 & 100\% \\
			& Proposed continuous & 20.38 & 112.49\% \\
			& Proposed discrete & 20.49 & 99.85\% \\
			\hline
		\end{tabular}
	\end{center}
\end{table}

This paper investigates the joint resource allocation and beamforming design problem in RIS-assisted MISO-OFDMA systems, aiming to maximize the system sum rate by optimizing RB allocation decisions, RIS reflection phase shift, and BS beamforming. To address the high computational complexity of traditional numerical optimization methods and the difficulty of obtaining labels for supervised learning approaches, a two-stage neural network based on UL is proposed. Specifically, the proposed scheme employs BeamNet and AllocationNet to output the RIS reflection phase shift and RB allocation decisions, respectively, followed by MRT and the water-filling algorithm to optimize the BS beamforming. To overcome the challenge of discrete output in neural networks, a quantization layer and the Gumbel-softmax trick are introduced to enable BeamNet and AllocationNet to output discrete RIS reflection phase shift and RB allocation decisions.
In addition, we propose a customized loss function to address the inequality constraints inherent to the optimization problem.
Simulation results demonstrate that the proposed approach achieves 99.93\% of the system sum rate of the SCA method while requiring only 0.036\% of its runtime.
Although the proposed method demonstrates strong performance and robustness under various scenarios, it still requires retraining when the number of subcarriers or the number of RIS reflecting elements changes. This limitation stems from the fixed input and output dimensions of the neural networks. As a direction for future work, we plan to explore scalable and flexible network architectures that can adapt to varying system dimensions without retraining. Additionally, meta-learning techniques could be investigated to enable rapid adaptation to new configurations.

\bibliographystyle{IEEEtran}

\bibliography{references.bib}

\begin{thebibliography}{10}
\providecommand{\url}[1]{#1}
\csname url@samestyle\endcsname
\providecommand{\newblock}{\relax}
\providecommand{\bibinfo}[2]{#2}
\providecommand{\BIBentrySTDinterwordspacing}{\spaceskip=0pt\relax}
\providecommand{\BIBentryALTinterwordstretchfactor}{4}
\providecommand{\BIBentryALTinterwordspacing}{\spaceskip=\fontdimen2\font plus
\BIBentryALTinterwordstretchfactor\fontdimen3\font minus
  \fontdimen4\font\relax}
\providecommand{\BIBforeignlanguage}[2]{{%
\expandafter\ifx\csname l@#1\endcsname\relax
\typeout{** WARNING: IEEEtran.bst: No hyphenation pattern has been}%
\typeout{** loaded for the language `#1'. Using the pattern for}%
\typeout{** the default language instead.}%
\else
\language=\csname l@#1\endcsname
\fi
#2}}
\providecommand{\BIBdecl}{\relax}
\BIBdecl

\bibitem{WirelessZhang}
Z.~Zhang \emph{et~al.}, ``{6G} wireless networks: {Vision}, requirements,
  architecture, and key technologies,'' \emph{IEEE Veh. Technol. Mag.},
  vol.~14, no.~3, pp. 28--41, 2nd Quart. 2019.

\bibitem{UltraMassive}
A.~M. Elbir, K.~V. Mishra, and S.~Chatzinotas, ``{Terahertz}-band joint
  ultra-massive {MIMO} radar-communications: {Model}-based and model-free
  hybrid beamforming,'' \emph{IEEE J. Sel. Top. Signal Process}, vol.~15,
  no.~6, pp. 1468--1483, Nov. 2021.

\bibitem{Terahertz}
I.~F. Akyildiz, C.~Han, Z.~Hu, S.~Nie, and J.~M. Jornet, ``Terahertz band
  communication: {An} old problem revisited and research directions for the
  next decade,'' \emph{IEEE Trans. Commun.}, vol.~70, no.~6, pp. 4250--4285,
  Jun. 2022.

\bibitem{ISAC1}
X.~Wang, Z.~Fei, and Q.~Wu, ``Integrated sensing and communication for
  {RIS}-assisted backscatter systems,'' \emph{IEEE Internet Things J.},
  vol.~10, no.~15, pp. 13\,716--13\,726, Aug. 2023.

\bibitem{ISAC2}
N.~Wu, X.~Wang, Z.~Fei, F.~Xia, J.~Huang, and A.~Nallanathan, ``{RIS}-assisted
  integrated sensing and backscatter communications for future {IoT}
  networks,'' \emph{IEEE Internet of Things Mag.}, vol.~7, no.~4, pp. 44--50,
  Jul. 2024.

\bibitem{reviewer2_1}
H.~Liu \emph{et~al.}, ``Stacked intelligent metasurfaces for wireless sensing
  and communication: {Applications} and challenges,'' \emph{arXiv preprint
  arXiv:2407.03566}, 2024.

\bibitem{reviewer2_2}
H.~Liu, J.~An, G.~C. Alexandropoulos, D.~W.~K. Ng, C.~Yuen, and L.~Gan,
  ``Multi-user {MISO} with stacked intelligent metasurfaces: {A} {DRL}-based
  sum-rate optimization approach,'' \emph{IEEE Trans. Cognit. Commun.
  Networking}, pp. 1--1, 2025.

\bibitem{SmartDi}
M.~Di~Renzo \emph{et~al.}, ``Smart radio environments empowered by
  reconfigurable intelligent surfaces: {How} it works, state of research, and
  the road ahead,'' \emph{IEEE J. Sel. Areas Commun.}, vol.~38, no.~11, pp.
  2450--2525, Nov. 2020.

\bibitem{ReconfigurablePan}
C.~Pan \emph{et~al.}, ``Reconfigurable intelligent surfaces for {6G} systems:
  {Principles}, applications, and research directions,'' \emph{IEEE Commun.
  Mag.}, vol.~59, no.~6, pp. 14--20, Jun. 2021.

\bibitem{9999288}
J.~Sang \emph{et~al.}, ``Coverage enhancement by deploying ris in {5G}
  commercial mobile networks: {Field} trials,'' \emph{IEEE Wireless Commun.},
  vol.~31, no.~1, pp. 172--180, Feb. 2024.

\bibitem{ReconfigurableLiu}
Y.~Liu \emph{et~al.}, ``Reconfigurable intelligent surfaces: {Principles} and
  opportunities,'' \emph{IEEE Commun. Surv.}, vol.~23, no.~3, pp. 1546--1577,
  3rd Quart. 2021.

\bibitem{single-user}
Q.~Wu and R.~Zhang, ``Intelligent reflecting surface enhanced wireless network:
  {Joint} active and passive beamforming design,'' in \emph{Proc. IEEE
  GLOBECOM}, Dec. 2018, pp. 1--6.

\bibitem{single-user1}
------, ``Intelligent reflecting surface enhanced wireless network via joint
  active and passive beamforming,'' \emph{IEEE Trans. Wireless Commun.},
  vol.~18, no.~11, pp. 5394--5409, Nov. 2019.

\bibitem{ADMM}
H.~Guo, Y.-C. Liang, J.~Chen, and E.~G. Larsson, ``Weighted sum-rate
  maximization for intelligent reflecting surface enhanced wireless networks,''
  in \emph{Proc. IEEE GLOBECOM}, Dec. 2019, pp. 1--6.

\bibitem{SDR}
Q.~Wu and R.~Zhang, ``Intelligent reflecting surface enhanced wireless network:
  {Joint} active and passive beamforming design,'' in \emph{Proc. IEEE
  GLOBECOM}, Dec. 2018, pp. 1--6.

\bibitem{discrete}
B.~Di, H.~Zhang, L.~Song, Y.~Li, Z.~Han, and H.~V. Poor, ``Hybrid beamforming
  for reconfigurable intelligent surface based multi-user communications:
  {Achievable} rates with limited discrete phase shifts,'' \emph{IEEE J. Sel.
  Areas Commun.}, vol.~38, no.~8, pp. 1809--1822, Aug. 2020.

\bibitem{discrete1}
H.~Gao, K.~Cui, C.~Huang, and C.~Yuen, ``Robust beamforming for {RIS}-assisted
  wireless communications with discrete phase shifts,'' \emph{IEEE Wireless
  Commun. Lett.}, vol.~10, no.~12, pp. 2619--2623, Dec. 2021.

\bibitem{discrete2}
J.~Sang \emph{et~al.}, ``Quantized phase alignment by discrete phase shifts for
  reconfigurable intelligent surface-assisted communication systems,''
  \emph{IEEE Trans. Veh. Technol.}, vol.~73, no.~4, pp. 5259--5275, Apr. 2024.

\bibitem{Rate}
S.~Hassouna, M.~A. Jamshed, M.~Ur-Rehman, K.~Arshad, M.~A. Imran, and Q.~H.
  Abbasi, ``Rate optimization and power allocation in {RIS}-assisted multi-user
  {OFDM} communication,'' in \emph{Proc. IEEE WCNC}, 2024, pp. 01--05.

\bibitem{power2}
K.~Feng, X.~Li, Y.~Han, and Y.~Chen, ``Joint beamforming optimization for
  reconfigurable intelligent surface-enabled {MISO-OFDM} systems,'' \emph{China
  Commun.}, vol.~18, no.~3, pp. 63--79, Mar 2021.

\bibitem{ofdmaLee}
J.~Lee, J.~Choi, and J.~Kang, ``Harmony search-based optimization for
  multi-{RISs} {MU}-{MISO} {OFDMA} systems,'' \emph{IEEE Wireless Commun.
  Lett.}, vol.~12, no.~2, pp. 257--261, Feb. 2023.

\bibitem{ofdmaWei}
Z.~Wei, Y.~Cai, Z.~Sun, D.~W.~K. Ng, J.~Yuan, M.~Zhou, and L.~Sun, ``Sum-rate
  maximization for {IRS}-assisted {UAV} {OFDMA} communication systems,''
  \emph{IEEE Trans. Wireless Commun.}, vol.~20, no.~4, pp. 2530--2550, Apr.
  2021.

\bibitem{Liang}
Z.~Qin, L.~Liang, Z.~Wang, S.~Jin, X.~Tao, W.~Tong, and G.~Y. Li, ``{AI}
  empowered wireless communications: {F}rom bits to semantics,'' \emph{Proc.
  IEEE}, vol. 112, no.~7, pp. 621--652, 2024.

\bibitem{channelestimation}
E.~Shtaiwi, H.~Zhang, A.~Abdelhadi, and Z.~Han, ``{RIS}-assisted mmwave channel
  estimation using convolutional neural networks,'' in \emph{Proc. IEEE WCNCW},
  2021, pp. 1--6.

\bibitem{supervised}
A.~Koc, M.~Wang, and T.~Le-Ngoc, ``Deep learning based multi-user power
  allocation and hybrid precoding in massive {MIMO} systems,'' in \emph{Proc.
  IEEE ICC}, 2022, pp. 5487--5492.

\bibitem{9090876}
A.~M. Elbir, A.~Papazafeiropoulos, P.~Kourtessis, and S.~Chatzinotas, ``Deep
  channel learning for large intelligent surfaces aided mm-{Wave} massive
  {MIMO} systems,'' \emph{IEEE Wireless Commun. Lett.}, vol.~9, no.~9, pp.
  1447--1451, Sep. 2020.

\bibitem{9370097}
A.~Taha, M.~Alrabeiah, and A.~Alkhateeb, ``{Enabling} large intelligent
  surfaces with compressive sensing and deep learning,'' \emph{IEEE Access},
  vol.~9, pp. 44\,304--44\,321, Mar. 2021.

\bibitem{label}
H.~Huang \emph{et~al.}, ``Deep learning for physical-layer {5G} wireless
  techniques: {Opportunities}, challenges and solutions,'' \emph{IEEE Wireless
  Commun.}, vol.~27, no.~1, pp. 214--222, Feb. 2020.

\bibitem{Unsupervised_Gao}
Z.~Liu, Y.~Yang, F.~Gao, T.~Zhou, and H.~Ma, ``Deep unsupervised learning for
  joint antenna selection and hybrid beamforming,'' \emph{IEEE Trans. Commun.},
  vol.~70, no.~3, pp. 1697--1710, Mar. 2022.

\bibitem{Semi-Supervised}
H.~Sifaou and O.~Simeone, ``Semi-supervised learning via
  cross-prediction-powered inference for wireless systems,'' \emph{IEEE trans.
  mach. learn. commun. netw.}, vol.~3, pp. 30--44, Nov. 2024.

\bibitem{Self-Supervised}
M.~Farahmandand and M.~Nabi, ``Channel quality prediction for {TSCH}
  blacklisting in highly dynamic networks: {A} self-supervised deep learning
  approach,'' \emph{IEEE Sens. J.}, vol.~21, no.~18, pp. 21\,059--21\,068, Sep.
  2021.

\bibitem{UnsupervisedJiabao}
J.~Gao, C.~Zhong, X.~Chen, H.~Lin, and Z.~Zhang, ``Unsupervised learning for
  passive beamforming,'' \emph{IEEE Commun. Lett.}, vol.~24, no.~5, pp.
  1052--1056, May 2020.

\bibitem{UnsupervisedSong}
H.~Song, M.~Zhang, J.~Gao, and C.~Zhong, ``Unsupervised learning-based joint
  active and passive beamforming design for reconfigurable intelligent surfaces
  aided wireless networks,'' \emph{IEEE Commun. Lett.}, vol.~25, no.~3, pp.
  892--896, Mar. 2021.

\bibitem{10533223}
J.~Ye, L.~Huang, Z.~Chen, P.~Zhang, and M.~Rihan, ``Unsupervised learning for
  joint beamforming design in {RIS}-aided {ISAC} systems,'' \emph{IEEE Wireless
  Commun. Lett.}, vol.~13, no.~8, pp. 2100--2104, 2024.

\bibitem{OFDMA_zhangRui}
Y.~Yang, S.~Zhang, and R.~Zhang, ``{IRS}-enhanced {OFDMA}: Joint resource
  allocation and passive beamforming optimization,'' \emph{IEEE Wireless
  Commun. Lett.}, vol.~9, no.~6, pp. 760--764, Jun. 2020.

\bibitem{singal}
P.~Chen, W.~Huang, X.~Li, and S.~Jin, ``Deep reinforcement learning based power
  minimization for {RIS}-assisted {MISO}-{OFDM} systems,'' \emph{China
  Commun.}, vol.~20, no.~4, pp. 259--269, Apr. 2023.

\bibitem{prefix}
Y.~Yang, S.~Zhang, and R.~Zhang, ``{IRS}-enhanced {OFDM}: {Power} allocation
  and passive array optimization,'' in \emph{Proc. IEEE GLOBECOM}, 2019, pp.
  1--6.

\bibitem{DFT}
Y.~Yang, B.~Zheng, S.~Zhang, and R.~Zhang, ``Intelligent reflecting surface
  meets {OFDM}: {Protocol} design and rate maximization,'' \emph{IEEE Trans.
  Commun.}, vol.~68, no.~7, pp. 4522--4535, Jul. 2020.

\bibitem{DRL}
Y.~Zhao, Y.~Kim, and J.~Lee, ``{SOQ}: {Structural} reinforcement learning for
  constrained delay minimization with channel state information,'' \emph{IEEE
  Internet of Things Journal}, vol.~11, no.~3, pp. 4628--4644, Feb. 2024.

\bibitem{DRL1}
L.~Liang, H.~Ye, and G.~Y. Li, ``Spectrum sharing in vehicular networks based
  on multi-agent reinforcement learning,'' \emph{IEEE J. Sel. Areas Commun.},
  vol.~37, no.~10, pp. 2282--2292, Oct. 2019.

\bibitem{jang2016categorical}
E.~Jang, S.~Gu, and B.~Poole, ``Categorical reparameterization with
  gumbel-softmax,'' \emph{arXiv preprint arXiv:1611.01144}, 2016.

\bibitem{TowardsLiang}
F.~Liang, C.~Shen, W.~Yu, and F.~Wu, ``Towards optimal power control via
  ensembling deep neural networks,'' \emph{IEEE Trans. Commun.}, vol.~68,
  no.~3, pp. 1760--1776, Mar. 2020.

\bibitem{gain}
E.~Basar, M.~Di~Renzo, J.~De~Rosny, M.~Debbah, M.-S. Alouini, and R.~Zhang,
  ``Wireless communications through reconfigurable intelligent surfaces,''
  \emph{IEEE Access}, vol.~7, pp. 116\,753--116\,773, Aug. 2019.

\bibitem{dis}
B.~Di, H.~Zhang, L.~Song, Y.~Li, Z.~Han, and H.~V. Poor, ``Hybrid beamforming
  for reconfigurable intelligent surface based multi-user communications:
  {Achievable} rates with limited discrete phase shifts,'' \emph{IEEE J. Sel.
  Areas Commun.}, vol.~38, no.~8, pp. 1809--1822, Aug. 2020.

\bibitem{LearningNaderiAlizadeh}
N.~NaderiAlizadeh, M.~Eisen, and A.~Ribeiro, ``Learning resilient radio
  resource management policies with graph neural networks,'' \emph{IEEE Trans.
  Signal Process.}, vol.~71, pp. 995--1009, 2023.

\end{thebibliography}

\end{document}